\documentclass[useAMS,usenatbib]{mn2e}
\pdfoutput=1

\usepackage{subfigure}
\usepackage{url}
\usepackage{deluxetable}
\usepackage{longtable}
\usepackage{rotating}
\usepackage{amssymb,amsmath}
\usepackage{natbib}  
\newcommand\apj {\rmfamily{ApJ}}   
\newcommand\apjs{\rmfamily{ApJS}}  
\newcommand\apjl{\rmfamily{ApJ}}   
\newcommand\aap {\rmfamily{A\&A}}  
\newcommand\araa{\rmfamily{ARA\&A}}  
\newcommand\mnras{\rmfamily{MNRAS}}
\newcommand{\ds}{\ensuremath{\textrm{d}s}}
\newcommand{\dv}{\ensuremath{\textrm{d}v}}
\newcommand{\dnu}{\ensuremath{\textrm{d}\nu}}
\newcommand{\msun}{\ensuremath{M_{\odot}}}			
\newcommand{\hh}{\ensuremath{\textrm{H}_{2}}}			
\newcommand{\kms}{\textrm{km~s}\ensuremath{^{-1}}}	
\newcommand{\percc}{\ensuremath{\textrm{cm}^{-3}}}
\newcommand{\persc}{\ensuremath{\textrm{cm}^{-2}}}
\newcommand{\persr}{\ensuremath{\textrm{sr}^{-1}}}
\newcommand{\peryr}{\ensuremath{\textrm{yr}^{-1}}}
\newcommand{\um}{\ensuremath{\mu m}}    

\newcommand{\twelveco}{\ensuremath{^{12}\textrm{CO}}}
\newcommand{\thirteenco}{\ensuremath{^{13}\textrm{CO}}}

\def\ee#1{\ensuremath{\times10^{#1}}}
\newcommand{\degree}{\ensuremath{^{\circ}}}
\newcommand{\lowirac}{800}
\newcommand{\highirac}{8000}

\newcommand{\nwfive}{40}
\newcommand{\nouter}{15}
\newcommand{\noutflows}{55}

\newcommand\epsscale[1]{\gdef\eps@scaling{#1}}
\newcommand\plotone[1]{%
 \typeout{Plotone included the file #1}
 \centering
 \leavevmode
 \includegraphics[width={\columnwidth}]{#1}%
}%

\def\Figure#1#2#3#4{
\begin{figure*}
\epsscale{#4}
\plotone{#1}
\caption{#2}
\label{#3}
\end{figure*}
}

\def\FigureTwo#1#2#3#4{
\begin{figure*}
\subfigure[]{ \includegraphics[width=4in]{#1} }
\subfigure[]{ \includegraphics[width=4in]{#2} }
\caption{#3}
\label{#4}
\end{figure*}
}

\def\FigureFour#1#2#3#4#5#6{
\begin{figure*}
\subfigure[]{ \includegraphics[width=3in]{#1} }
\subfigure[]{ \includegraphics[width=3in]{#2} }
\subfigure[]{ \includegraphics[width=3in]{#3} }
\subfigure[]{ \includegraphics[width=3in]{#4} }
\caption{#5}
\label{#6}
\end{figure*}
}

\def\FigureFour#1#2#3#4#5#6{
\begin{figure*}
\subfigure[]{ \includegraphics[width=3in]{#1} }
\subfigure[]{ \includegraphics[width=3in]{#2} }
\subfigure[]{ \includegraphics[width=3in]{#3} }
\subfigure[]{ \includegraphics[width=3in]{#4} }
\caption{#5}
\label{#6}
\end{figure*}
}

\def\Table#1#2#3#4#5#6{
\begin{table*}
\caption{#2}
\label{#4}
\begin{tabular}{#1}
#3
\hline
#5
\hline
\end{tabular}
\footnote{#6}
\end{table*}
}

\def\LongTable#1#2#3#4#5#6#7#8{
\renewcommand{\thefootnote}{\alph{footnote}}
\begin{longtable}{#1}
\caption[#2]{#2}
\label{#4} \\

 \\
\hline 
#3 \\
\hline
\endfirsthead

\hline
#3 \\
\hline
\endhead

\hline
\multicolumn{#8}{r}{{Continued on next page}} \\
\hline
\endfoot

\hline 
\endlastfoot
#7 \\

#5
\hline
#6 \\

\end{longtable}
\renewcommand{\thefootnote}{\arabic{footnote}}
}

\begin{document}

\title{JCMT HARP CO 3-2 Observations of Molecular Outflows in W5}

\author[A. Ginsburg and J. Bally and J.P. Williams]{Adam
Ginsburg$^1$\thanks{adam.ginsburg@colorado.edu}, John Bally$^1$,
and Jonathan P. Williams$^2$\\
$^1$Center for Astrophysics and Space Astronomy, 
University of Colorado
389 UCB, Boulder, CO 80309-0389\\
$^2$Institute for Astronomy
University of Hawaii
2680 Woodlawn Dr. Honolulu, HI 96822}

\maketitle

\begin{abstract}

New JCMT HARP CO 3-2 observations of the W5 star forming complex are presented,
totaling an area of $\sim12000$ arcmin$^2$ with sensitivity better than 0.1 K
per 0.4 \kms\ channel.  We discovered \noutflows\ CO outflow candidates, of which \nwfive\ are
associated with W5 and \nouter\ are more distant than the Perseus arm.  Most of
the outflows are located on the periphery of the W5 HII region. However, two
outflow clusters are $>5$ pc from the ionization fronts, indicating that their
driving protostars formed without directly being triggered by the O-stars in
W5.  We compare the derived outflow properties to those in Perseus and find
that the total W5 outflow mass is surprisingly low given the cloud masses.  The
outflow mass deficiency in the more massive W5 cloud ($M(\hh)\sim5\ee{4}\msun$)
can be explained if ionizing radiation dissociates   molecules as they break
out of their host cloud cores.  Although CO J=3-2 is a good outflow tracer, it
is likely to be a poor mass tracer because of sub-thermal line excitation and
high opacity, which may also contribute to the outflow mass discrepancy.
It is unlikely
that outflows could provide the observed turbulent energy in the W5 molecular
clouds even accounting for undetected outflow material.  Many cometary globules
have been observed with velocity gradients from head to tail, displaying strong
interaction with the W5 HII region and exhibiting signs of triggered or revealed
star formation in their heads.  Because it is observed face-on, W5 is an
excellent region to study feedback effects, both positive and negative, of
massive stars on star formation.

\end{abstract}

\begin{keywords}ISM: jets and outflows ---
          ISM: kinematics and dynamics ---
          ISM: individual: W5 ---
          stars: formation
\end{keywords}

\section{Introduction}

Galactic-scale shocks such as spiral density waves promote the formation of
giant molecular clouds (GMCs) where massive stars, star clusters, and OB
associations form.  The massive stars in such groups can either disrupt the
surrounding medium or promote further star formation.  While ionizing and soft
UV radiation, stellar winds, and eventually supernova explosions destroy clouds
in the immediate vicinity of massive stars, as the resulting bubbles age and
decelerate, they can also trigger further star formation.  In the ``collect and
collapse'' scenario
\citep[e.g.][]{elmegreen:sequential:1977}, gas swept-up by expanding bubbles
can collapse into new star-forming clouds.  In the ``radiation-driven
implosion'' model \citep{bertoldi:cometary:1990,klein:implosion:1983},
pre-existing clouds may be compressed by photo-ablation pressure or by the
increased pressure as they are overrun by an expanding shell.  In some
circumstances, forming stars are simply exposed as low-density gas is removed
by winds and radiation from massive stars.  These processes may play significant roles in
determining the efficiency of star formation in clustered environments
\citep{elmegreen1998}.

Feedback from low mass stars may also control the shape of the stellar initial
mass function in clusters \citep{adams1996,peters2010}.  Low mass young stars
generate high velocity, collimated outflows that contribute to the turbulent
support of a gas clump, preventing the clump from forming stars long enough
that it is eventually blown away by massive star feedback.  It is therefore
important to understand the strength of low-mass protostellar feedback relative
to other feedback mechanisms.

Outflows are a ubiquitous indicator of the presence of ongoing star formation
\citep{reipurth2001}.  CO outflows are an indicator of ongoing embedded star
formation at a younger stage than optical outflows because shielding from the
interstellar radiation field is required for CO to survive.  Although Herbig-
Haro shocks and \hh\ knots reveal the locations of the highest-velocity
segments of these outflows, CO has typically been thought of as a
``calorimeter'' measuring the majority of the mass and momentum ejected from
protostars or swept up by the ejecta \citep{Bachiller1996}.

\begin{figure*}
  \includegraphics[angle=90,width=5in]{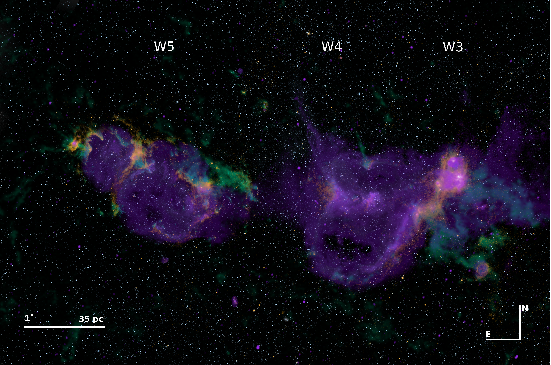}
  \caption{An overview of the W3/4/5 complex (also known as the ``Heart and
  Soul Nebula'') in false color. Orange shows 8 \um\ emission from the Spitzer
  and MSX satellites.  Purple shows 21 cm continuum emission from the DRAO CGPS
  \citep{Taylor2003:CGPS}; the DSS R image was used to set the display opacity
  of the 21 cm continuum as displayed (purely for aesthetic purposes).  The
  green shows JCMT \twelveco\ 3-2 along with FCRAO \twelveco\ 1-0 to fill in 
  gaps that were not observed with the JCMT.  The image spans
  $\sim7\degree$ in galactic longitude.  This overview image shows the
  hypothesized interaction between the W4 superbubble and the W3 and W5
  star-forming regions \citep{oey:hierarchical:2005}.}
  \label{fig:color_overview}
\end{figure*}

The W5 star forming complex in the outer galaxy is a prime location to study
massive star formation and triggering.  The bright-rimmed clouds in W5 have
been recognized as good candidates for ongoing triggering by a number of groups
\citep{lefloch:cometary:1997,thompson:searching:2004,karr:triggered:2003}.  The
clustering properties were analyzed by \citet{koenig:clustered:2008} using
Spitzer infrared data, and a number of significant clusters were discovered.
The whole W5 complex may be a product of triggering, as it is located on one
side of the W4 chimney thought to be created by multiple supernovae during the
last $\sim$10 MYr \citep[][Figure \ref{fig:color_overview}]
{oey:hierarchical:2005}.  

Following \citet{koenig:clustered:2008}, we adopt a distance to W5 of 2 kpc
based on the water-maser parallax distance to the neighboring W3(OH) region
\citep{Hachisuka2006}.  As with W3, the W5 cloud is substantially
($\approx1.5\times$) closer than its kinematic distance would suggest
($v_{LSR}(-40~\kms)\approx3$ kpc).  Given this distance,
\citet{koenig:clustered:2008} derived a total gas mass of 6.5\ee{4} \msun\ from
a 2 \um\ extinction map.

The W5 complex was mapped in the $^{12}$CO 1-0 emission line by the Five
College Radio Astronomy Observatory (FCRAO) using the SEQUOIA receiver array
\citep{heyer:ogs:1998}.  The same array was used to map W5 in the \thirteenco\
1-0 line (C. Brunt, private communication).  Some early work searched for
outflows in W5 \citep{bretherton:unbiased:2002}, but the low-resolution CO 1-0
data only revealed a few, and only one was published.  The higher resolution
and sensitivity observations presented here reveal many additional outflows.

While W5 is thought to be associated with the W3/4/5 complex, there are other
infrared sources in the same part of the sky that are not obviously associated
with W5.  Some of these have been noted to be in the outer arm (several kpc
behind W5) by \citet{Digel1996} and \citet{Snell2002}.

\par
\par In section 2, we present the new and archival data used in our study.  In
section 3, we discuss the outflow detection process and compare outflow
detectability in W5 to that in Perseus.  In section 4, we discuss the physical
properties of the outflows and their implications for star formation in the W5
complex.  In section 5, we briefly describe the outer-arm outflows discovered.

\section{OBSERVATIONS}
\subsection{JCMT HARP CO 3-2}
CO J=3-2 345.79599 GHz data were acquired at the 15 m James Clerk Maxwell
Telescope (JCMT) using the HARP array on a series of observing runs in 2008.
On 2-4 January, 2008, $\sim$ 800 square arcminutes were mapped.  During the
run, $\tau_{225}$, the zenith opacity at 225 GHz measured using the Caltech
Submillimeter Observator tipping radiometer, ranged from 0.1 to 0.4
($0.4<\tau_{345
GHz}<1.6$\footnote{\url{http://docs.jach.hawaii.edu/JCMT/SCD/SN/002.2/node5.html}}).
Additional areas were mapped on 4-7 August, 16-20 and 31 October, and 1 and
12-15 Nov 2008 in similar conditions.  A total of $\sim$ 3 square degrees (12000
arcmin$^2$) in the W5 complex were mapped (a velocity-integrated mosaic is
shown in Figure \ref{fig:outflows_on_co32}).

HARP is a 16 pixel SIS receiver array acting as a front-end to the ACSIS
digital auto-correlation spectrometer.  In January 2008, 14 of the 16 detectors
were functional.  In the 2nd half of 2008, 12 of 16 were functional,
necessitating longer scans to achieve similar S/N.

In January 2008, a single spectral window centered at 345.7959899 with bandwidth 1.0
GHz and channel width 488 kHz (0.42 \kms) was used.  In August 2008 and later, we used 250 MHz
bandwidth and 61 kHz (0.05 \kms) channel width.  At this frequency, the beam
FWHM is 14\arcsec\ (0.14 pc at a distance of 2 kpc)
\footnote{\url{http://docs.jach.hawaii.edu/JCMT/OVERVIEW/tel_overview/}}.

A raster mapping strategy was used.  In 2008, the array was shifted by 1/2 of
an array spacing (58.2\arcsec) between scans.  Data was sampled at a rate of
$0.6 s$ per integration.  Two perpendicular scans were used for each field
observed.  Most fields were 10$\times$10\arcmin\ and took $\sim45$ minutes.
When only 12 receptors were available, 1/4 array stepping (29.1\arcsec) was
used with a sample rate of $0.4 s$ per integration.

Data were reduced using the SMURF package within the STARLINK software distribution
\footnote{\url{http://starlink.jach.hawaii.edu/}}.  The SMURF command {\sc makecube} was used to
generate mosaics of contiguous sub-fields.  The data were gridded on to cubes
with 6\arcsec\ pixels and smoothed with a $\sigma=2$-pixel gaussian, resulting
in a map FWHM resolution of 18\arcsec (0.17 pc).  A linear fit was subtracted from each
spectrum over emission-free velocities (generally -60 to -50 and -20 to -10
\kms) to remove the baseline.  The final map RMS was $\sigma_{T_A^*}\sim
0.06-0.11 K$ in 0.42 \kms\ channels.

The sky reference position (off position) in January 2008 was J2000 2:31:04.069 +62:59:13.81.
In later epochs, off positions closer to the target fields were selected from blank sky regions
identified in January 2008 in order to increase observing efficiency.  A
main-beam efficiency $\eta_{mb}=0.60$ was used as per the JCMT website to
convert measurements to $T_{mb}$, though maps and spectra are presented in the
original $T_A^*$ units.

\subsection{FCRAO Outer Galaxy Survey}
The FCRAO Outer Galaxy Survey (OGS)
observed the W5 complex in \twelveco\  \citep{heyer:ogs:1998} and \thirteenco\
1-0 (C. Brunt, private communication).  The \thirteenco\ data cube achieved a
mean sensitivity of 0.35 K per 0.13 \kms\ channel, or 0.6 K \kms\ integrated.
The \thirteenco\ cube was integrated over all velocities and resampled to match
the BGPS map using the {\sc montage}\footnote{\url{http://montage.ipac.caltech.edu/}}
package.  The FWHM beam size was  $\theta_{B}=$50\farcs (0.48 pc).  The integrated
\twelveco\ data cube, with a sensitivity $\sigma= 1 K$ \kms, is displayed with 
region name identifications in Figure \ref{fig:regionboxes_on_CO}.

\subsection{Spitzer}
Spitzer IRAC and MIPS 24 \um\ images from \citet{koenig:clustered:2008} were
used for morphological comparison.  The reduction and extraction techniques are
detailed in their paper.

\begin{figure*}
  \includegraphics[angle=90,width=5in]{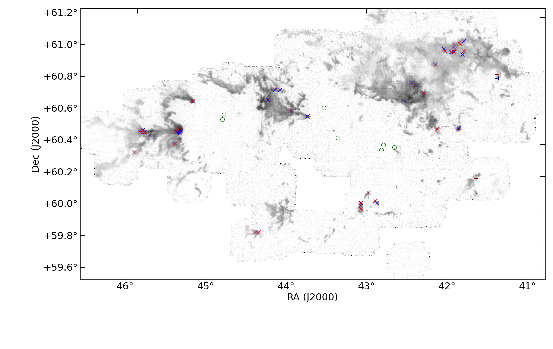}
  \caption{A mosaic of the CO 3-2 data cube integrated from -20 to -60 \kms.
  The grayscale is linear from 0 to 150 K \kms.  The red and blue X's mark the
  locations of redshifted and blueshifted outflows.  Dark red and dark blue
  plus symbols mark outflows at outer arm velocities.  Green circles mark the
  location of all known B0 and earlier stars in the W5 region from SIMBAD.}
  \label{fig:outflows_on_co32}
\end{figure*}

\Figure{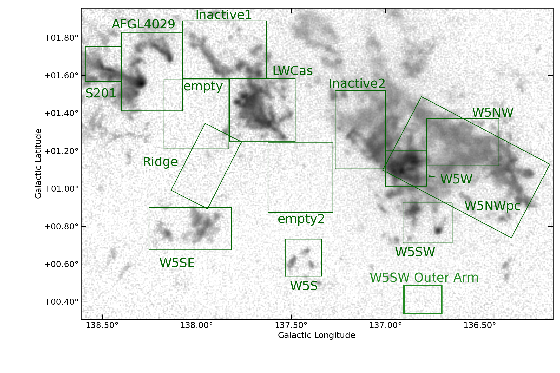}
{Individual region masks overlaid on the FCRAO \twelveco\ integrated image.
The named regions, S201, AFGL4029, LWCas, W5NW, W5W, W5SE, W5S, and W5SW, were all
selected based on the presence of outflows within the box.
The inactive regions were selected from regions with substantial CO emission
but without outflows.  The `empty' regions have essentially no CO emission within
them and are used to place limits on the molecular gas within the east and west
`bubbles'.  W5NWpc is compared directly to the Perseus molecular cloud in 
Section \ref{sec:percompare}
}{fig:regionboxes_on_CO}{1.0}

\section{Analysis}
\subsection{Outflow Detections}
Outflows were identified in the CO data cubes by manually searching through
position-velocity space for line wings using STARLINK's GAIA display software.  Outflow
candidates were identified by high velocity wings inconsistent with the local
cloud velocity distribution, which ranged from a width of 3 \kms\ to  7 \kms.
Once an outflow candidate was identified in the position-velocity diagrams, the
velocity range over which the wing showed emission in the position-velocity
diagram (down to $T_A^*=0$) was integrated over to create a map from which the
approximate outflow size and position was determined (e.g. Figures
\ref{fig:outflow1} and \ref{fig:pv2b}). 

Unlike \citet{curtis2010} and \citet{hatchell2009}, we did not use an
`objective' outflow identification method because of the greater velocity complexity
and poorer spatial resolution of our observations.  The outflow selection
criteria in these papers requires the presence of a sub-mm clump in order to
identify a candidate driving source (and therefore a targeted region in which to search for
outflows), making a similar objective identification impossible for
our survey.
As discussed later in 
Section \ref{sec:subregions}, the regions associated with outflows have wide
lines and many are double-peaked.  Additionally, many smaller areas associated
with outflows have collections of gaussian-profiled clumps that are not connected to
the cloud in position-velocity diagrams but are not outflows.  In particular,
W5 is pockmarked by dozens of small cometary globules that are sometimes
spatially coincident with the clouds but slightly offset in velocity.

While \citet{arce2010} described the benefits of 3D visualization using
isosurface contours, we found that the varying signal-to-noise across
large-scale ($\sim500$ pixel$^2$) regions with significant extent in RA/Dec and
limited velocity dynamic range made this method diffult for W5.  There
were many low-intensity outflows that were detectable by careful searches
through position-velocity space that are not as apparent using isosurface
methods.  Out of the 55 outflows reported here, only 14 \footnote{
Outflows 15, 20, 24, the cluster of outflows 26-32, 47, 48, 52, and 53 could 
all have readily been detected by pointed single-dish measurements.} would be
considered obvious, high-intensity, high-velocity flows from their spectra
alone; the rest could not be unambiguously detected without a search through
position-velocity space.

In the majority of sources, the individual outflow lobes were
unresolved, although some showed hints of position-velocity gradients at low
significance and in many the red and blue flows are spatially separated.    Only
Outflow 1's lobes were clearly resolved (Figure \ref{fig:outflow1}).  Some of
most suggestive gradients occurred where the outflow merged with its host
molecular cloud in position-velocity space, making the gradient difficult to
distinguish (e.g., Outflow 12, Figure \ref{fig:pv12}).  Bipolar pairs were
selected when there were red and blue flows close to one another.  The
classification of a bipolar flow was either `yc' (yes - confident), `yu' (yes -
unconfident), or `n' (no) in Table \ref{tab:outflows}.  This identification is
discussed in the captions for each outflow figure in the online
supplement. The AFGL 4029 region has many red and blue lobes but confusion
prevented pairing.  

\FigureFour{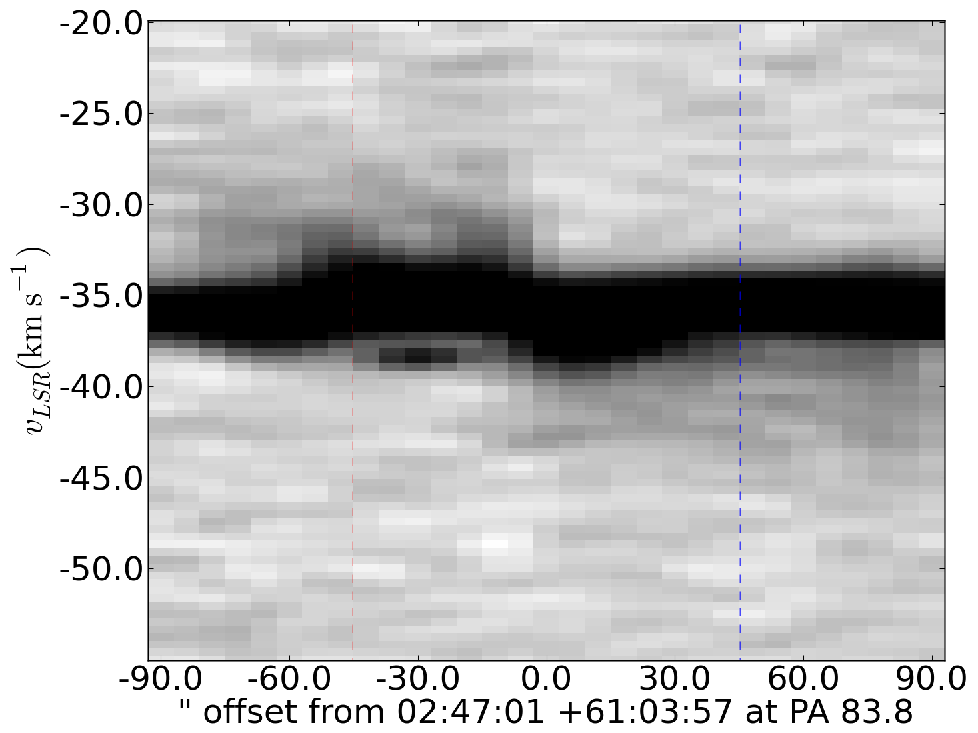}{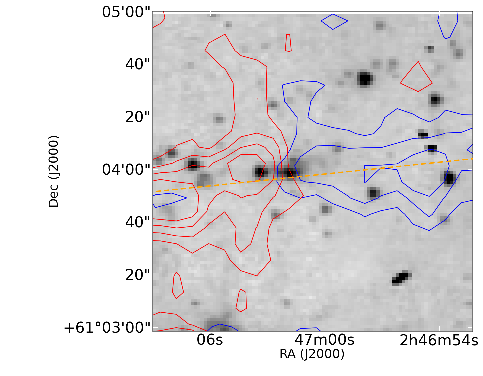}{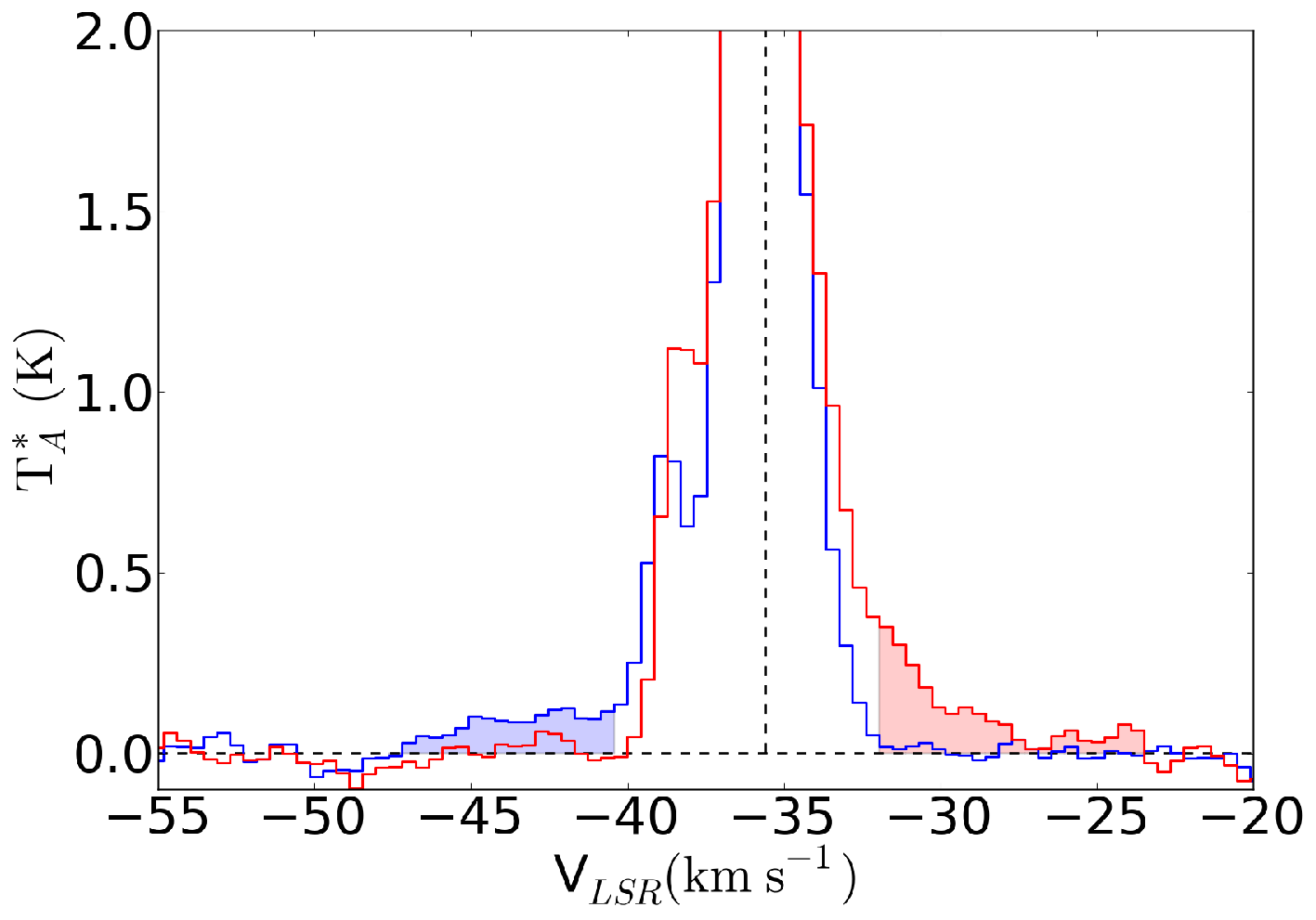}{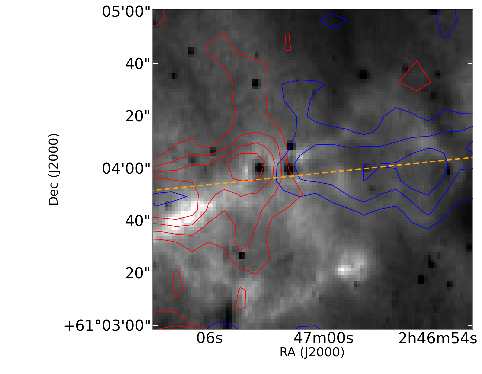}
{Position-velocity diagrams (a), spectra (c), and contour overlays of Outflow 1 on Spitzer 4.5 \um\ (b)
and 8 \um\ (d) images.  
This outflow is clearly resolved and bipolar.
{\it (a)}: Position-velocity diagram of the blue flow displayed in arcsinh stretch
from $T_A^*=$0 to 3 K.  Locations of the red and blue flows are indicated by vertical dashed lines.
The location of the position-velocity cut is indicated by the orange dashed line in panels (b) and
(d), although the position-velocity cut is longer than those cut-out images.
{\it (b)} Spitzer 4.5 \um\ image displayed in logarithmic stretch from 30 to 500 MJy \persr.    
{\it (c)}: Spectrum of
the outflow integrated over the outflow aperture and the velocity range
specified with shading.  The velocity center (vertical dashed line) is
determined by fitting a gaussian to the \thirteenco\ spectrum in an aperture
including both outflow lobes.  In the few cases in which \thirteenco\ 1-0 was 
unavailable, a gaussian was fit to the \twelveco\ 3-2 spectrum.
{\it (d)}: Contours of the red and blue outflows superposed on the 
Spitzer 8 \um\ image displayed in logarithmic stretch.  The contours are
generated from a total intensity image integrated over the outflow
velocities indicated in panel (c).  The contours in both panels (b) and (d) are displayed at levels of
0.5,1,1.5,2,3,4,5,6 K \kms\ ($\sigma\approx0.25$ K \kms).
The contour levels and stretches specified in this caption apply to all of the
figures in the supplementary materials except where otherwise noted.
}
{fig:outflow1}

\FigureFour{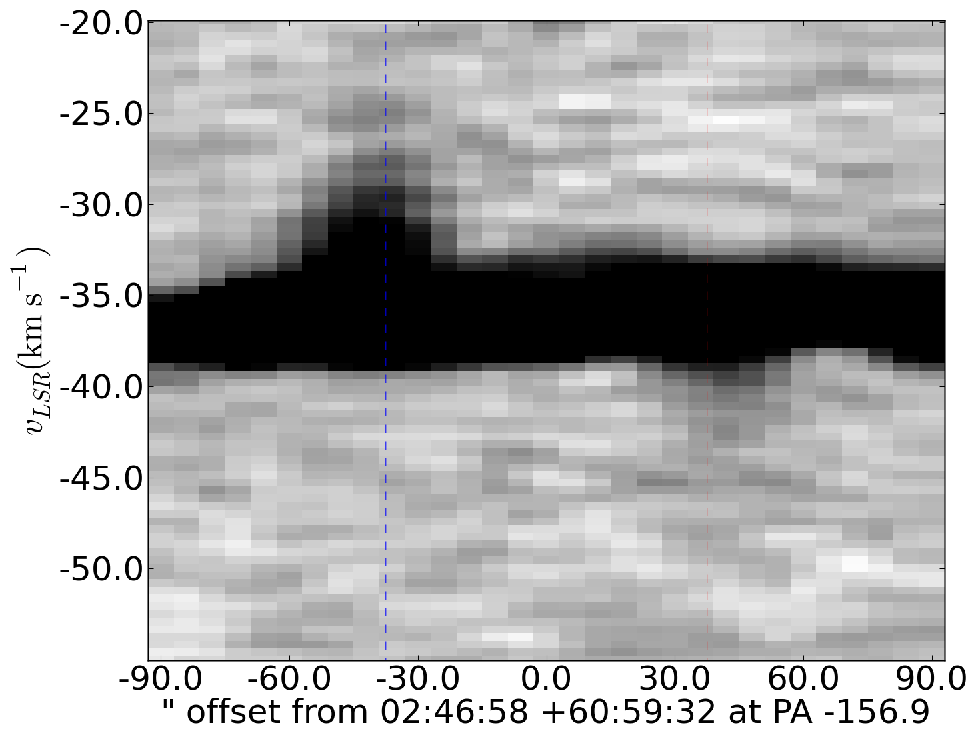}{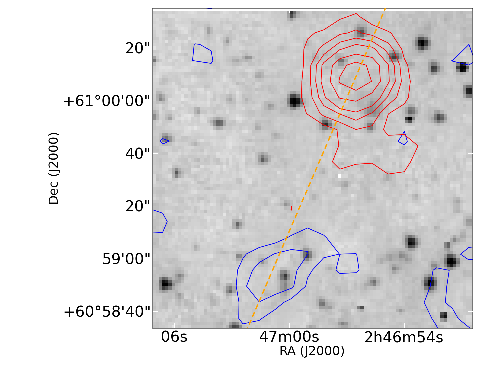}{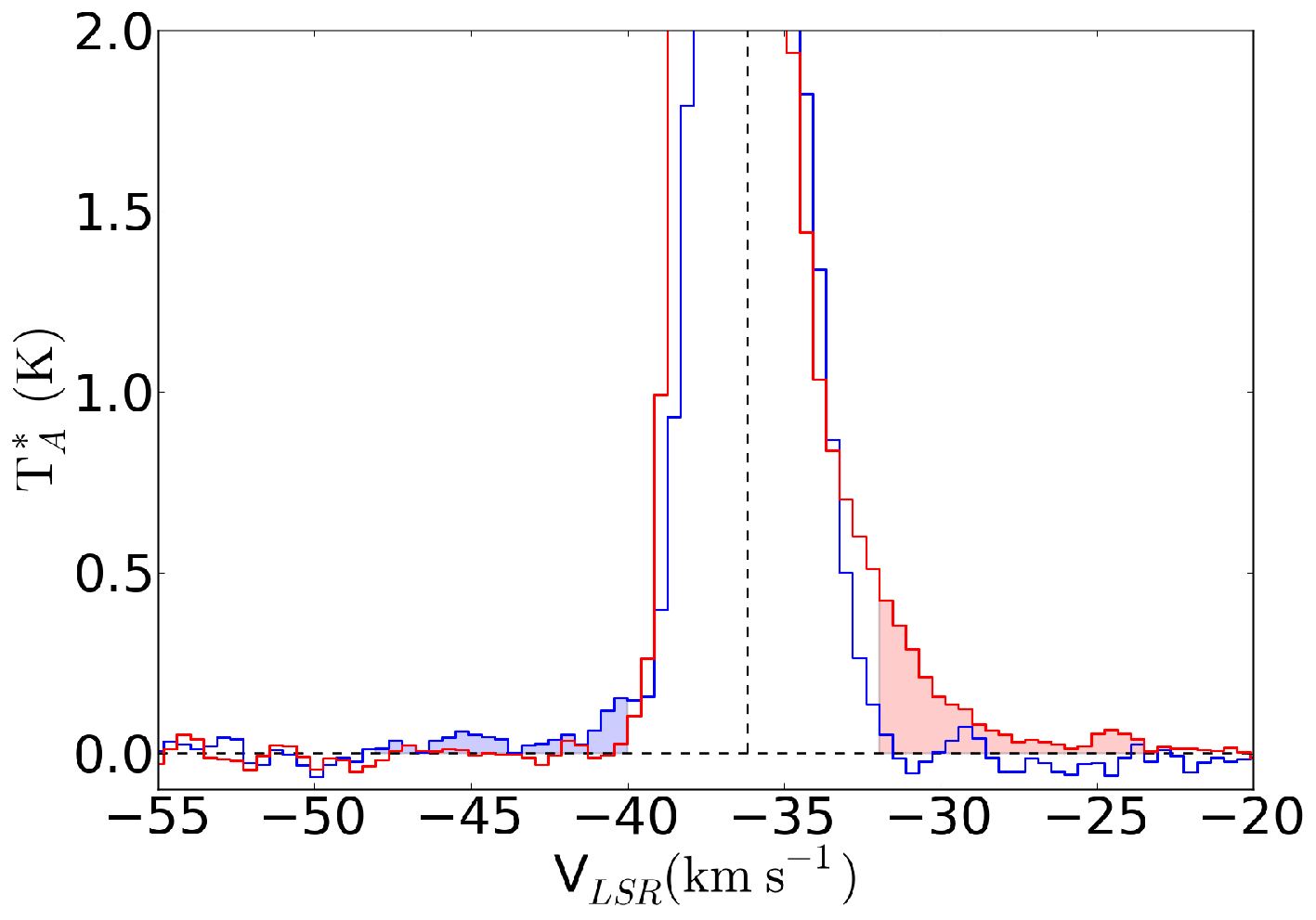}{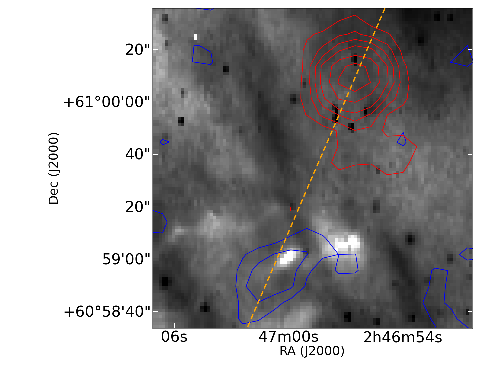}
{Position-velocity diagram, spectra, and contour overlays of Outflow 2 (see
Figure \ref{fig:outflow1} for a complete description).  While the two lobes are
widely separated, there are no nearby lobes that could lead to confusion, so we
regard this pair as a reliable bipolar outflow identification.  
}
{fig:pv2b}

\FigureFour{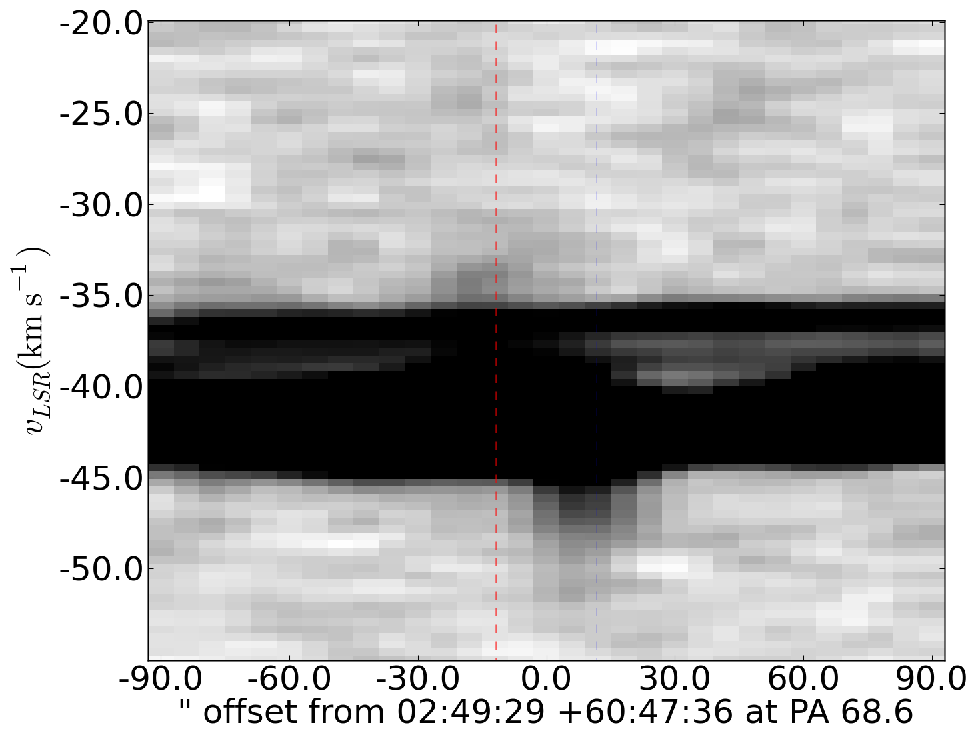}{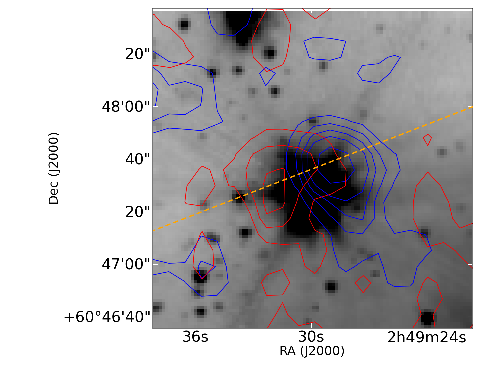}{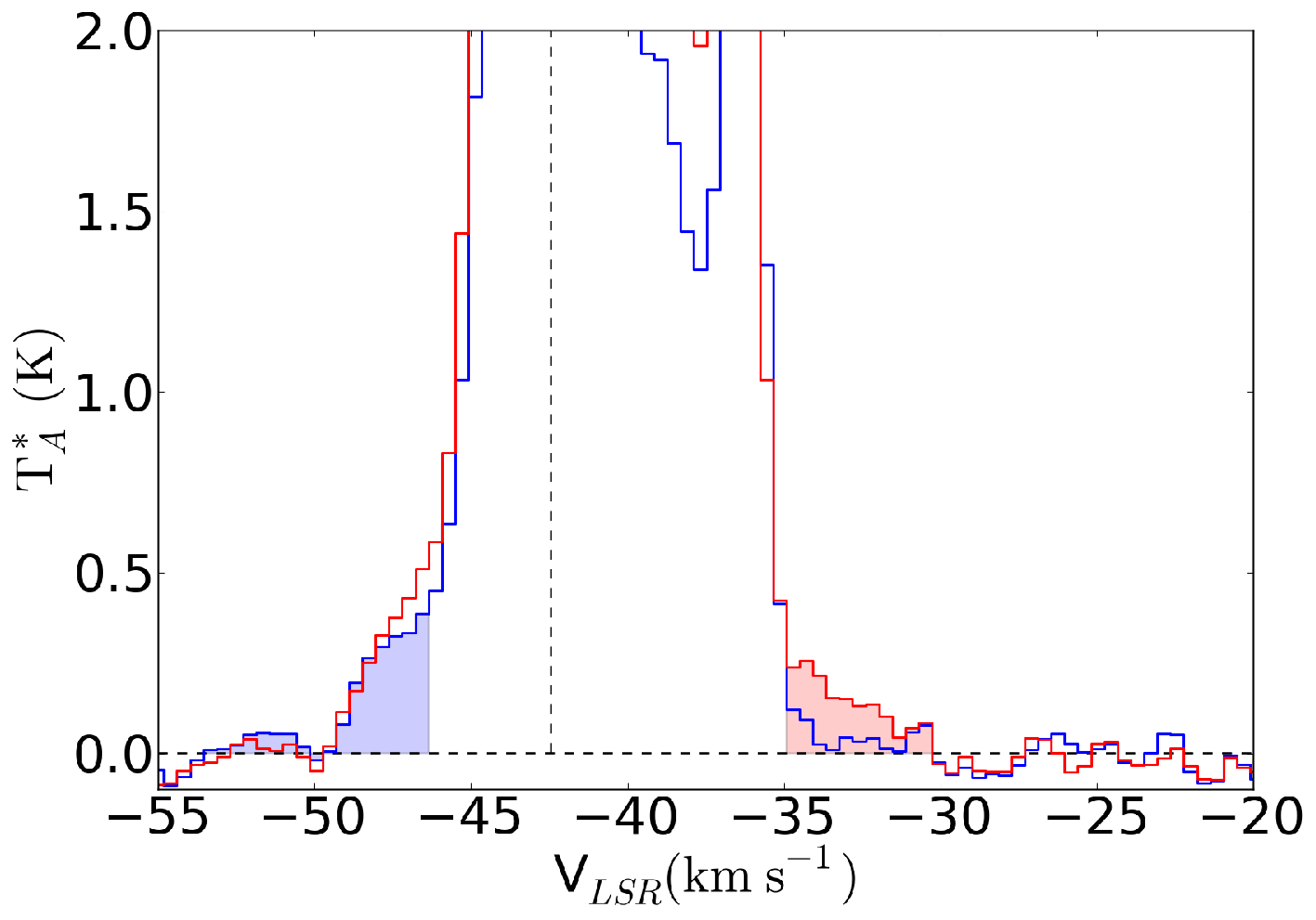}{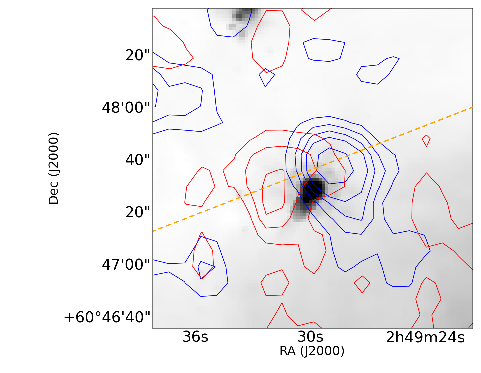}
{Position-velocity diagram, spectra, and contour overlays of Outflow 12.  Much of the red outflow
is lost in the complex velocity profile of the molecular cloud(s), but it is high enough velocity
to still be distinguished. }
{fig:pv12}

In cases where only the red- or blue-shifted lobe was visible, the
surrounding pixels were searched for lower-significance and lower-velocity
counterparts.  For cases in which emission was detected, a
candidate counterflow was identified and incorporated into the catalog.
However, in 12 cases, the counterflow still evaded detection, either because of
confusion or because the counterflow is not present in CO.

The outflow positions are overlaid on the CO 3-2 image in Figure
\ref{fig:outflows_on_co32} to provide an overview of where star formation is
most active.  The figures in Section \ref{sec:sfactivity}  show outflow
locations overlaid on small-scale images.

Because our detection method involved searching for high-velocity outflows
by eye, there should be no false detections.  However, it is possible
that some of these outflows are generated by mechanisms other than
protostellar jets and winds since we have not identified their driving sources.

One possible alternative driving mechanism is a photoevaporation flow,
which could be accelerated up to the sound speed of the ionized medium,
$c_{II} \approx 10$~\kms.
Gas accelerating away from the cloud would not be detected as an outflow
because it would be rapidly ionized.
However, gas driven inward would be accelerated and remain molecular.  It
could exhibit red and / or blue flows depending on the line of
sight orientation.  While there are viable candidates for this form of outflow
impersonator, such flows can only have peak velocities $v\lesssim c_{II}/4
\approx 2.5$~\kms\ in the strong adiabatic shock limit, so that any gas
seen with higher velocity tails are unlikely to be radiation-driven.  

Another plausible outflow impostor is the high-velocity tail in a
turbulent distribution.  However, for a typical molecular cloud, the low
temperatures would require very high mach-number shocks ($\mathcal{M}\gtrsim10$
assuming $T_{cloud}\sim20 $ K and $v_{flow} \sim 3$ \kms) that in idealized
turbulence should be rare and short-lived.  It is not known how frequent such
high-velocity excursions will be in non-ideal turbulence with gravity (A.
Goodman, P. Padoan, private communication).  Finally, it is less likely for
turbulent intermittency to have nearly coincident red and blue lobes, so
intermittency can be morphologically excluded in most cases.

\subsubsection{Comparison to Perseus CO 3-2 observations}  
\label{sec:percompare}
We used the HARP CO 3-2 cubes from \citet{hatchell2007} to evaluate our ability
to identify outflows.  We selected an outflow that was
well-resolved and unconfused, L1448, and evaluated it at both the native
sensitivity of the \citet{hatchell2007} observations and degraded in resolution
and sensitivity to match our own.  We focus on L1448 IRS2,
labeled Outflow 30 in \citet{hatchell2007}. Figure \ref{fig:l1448} shows a
comparison between the original quality and degraded data.  

Integrating over the outflow velocity range, we measure each lobe to be
about $1.6\arcmin\times0.8\arcmin$ ($0.14\times0.07$pc).  Assuming a distance
to Perseus of 250 pc \citep[e.g.][]{Enoch2006},  we smooth by a factor of 8 by convolving
the cube with a FWHM = 111\arcsec\ gaussian, then downsample by the same factor of 8
to achieve 6\arcsec\ square pixels at 2 kpc.  The resulting noise was reduced
because of the spatial and spectral smoothing and was measured to be $\approx 0.05$ K in
0.54 \kms\ channels, which is comparable to the sensitivity in our
survey.  It is still possible to distinguish the outflows
from the cloud in position-velocity space.  Each lobe is individually
unresolved (long axis $\sim12\arcsec$ compared to our beam FWHM of 18\arcsec),
but the two are separated by $\gtrsim 20\arcsec$ and therefore an overall
spatial separation can still be measured.  Because they are just barely
unresolved at this distance, the lobes' surface brightnesses are approximately
the same at 2 kpc as at 250 pc; if this outflow were seen at a greater distance
it would appear fainter.

\citet{hatchell2007} detected 4 outflows within this map, plus an additional
confused candidate.  We note an additional grouping of outflowing material in
the north-middle of the map(centered on coordinate 150$\times$150 in Figure
\ref{fig:l1448}).  In the smoothed version, only three outflows are detected in
the blue and two in the red, making flow-counterflow association difficult.
The north-central blueshifted component appears to be the counterpart of the
red flow when smoothed, although it is clearly the counterpart of the northwest
blue flow in the full-resolution image.

We are therefore able to detect any outflows comparable to L1448 (assuming a
favorable geometry), but are likely to see clustered outflows as single or
possibly extended lobes and will count fewer lobes than would be detected at
higher resolution.  Additionally, it is clear from this example that two
adjacent outflows with opposite polarity are not necessarily associated, and
therefore the outflows' source(s) may not be between the two lobes.  

\Figure{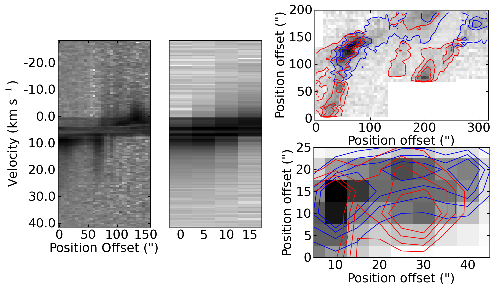}
{Comparison of L1448 seen at a distance of 250 pc (left) versus 2 kpc (middle) with
sensitivity 0.5 K and 0.05K per 0.5 \kms\ channel respectively.  {\it Far
Left}: Position-velocity diagram (log scale) of the outflow L1448 IRS2 at its
native resolution and velocity.  L1448 IRS2 is the rightmost outflow in the contour 
plots.  The PV diagram is rotated 45\degree\ from RA/Dec axes to go along the outflow
axis.
{\it Middle Left}: Position-velocity diagram (log scale) of the same outflow smoothed
and rebinned to be eight times more distant.
{\it Top Right}: The integrated map is displayed at its native resolution (linear scale).
The red contours are of the same data integrated from 6.5 to 16 \kms\ and the blue from 
-6 to 0 \kms.  Contours are at 1,3, and 5 K \kms\ ($\sim 6, 18, 30 \sigma$).  Axes
are offsets in arcseconds.  Because we are only examining the relative detectability of
outflows at two distances, we are not concerned with absolute coordinates.
{\it Bottom Right}: The same map as it would be observed at eight times greater
distance.  Axes are offsets in arcseconds assuming the greater distance.
Contours are integrated over the same velocity range as above, but are
displayed at levels 0.25,0.50,0.75,1.00 K \kms\ ($\sim 12, 24, 48, 60 \sigma$).
The entire region is detected at high significance, but dominated by confusion.
It is still evident that the red and blue lobes are distinct, but they are each
unresolved. 
}
{fig:l1448}{1.0}

In order to determine overall detectability of outflows compared to Perseus, we
compare to \citet{curtis2010} in Figure \ref{fig:lengthhist}.  Out of 29
outflows in their survey with measured `lobe lengths', 22 (71\%) were smaller
than 128\arcsec\, which would be below our 18\arcsec\ resolution if 
observed at 2 kpc.  Even the largest lobes (HRF26R,
HRF28R, HRF44B) would only extend $\sim60\arcsec$ at 2 kpc.  Each lobe in the
largest outflow in our survey, Outflow 1, is $\sim80\arcsec$ (660\arcsec\ at
250pc), but no other individual outflow lobes in W5 are clearly resolved.
However, as seen in Figure \ref{fig:lengthhist}, many bipolar lobes are
\emph{separated} by more than the telescope resolution, and the overall lobe
separation distribution (as opposed to the lobe length, which is mostly unmeasured
in our sample) in W5 is quite similar to the separation distribution in
Perseus.  The 2-sample KS test gives a 25\% probability that they are drawn
from the same distribution (the null hypothesis that they are drawn from the
same distribution cannot be rejected).  

\Figure{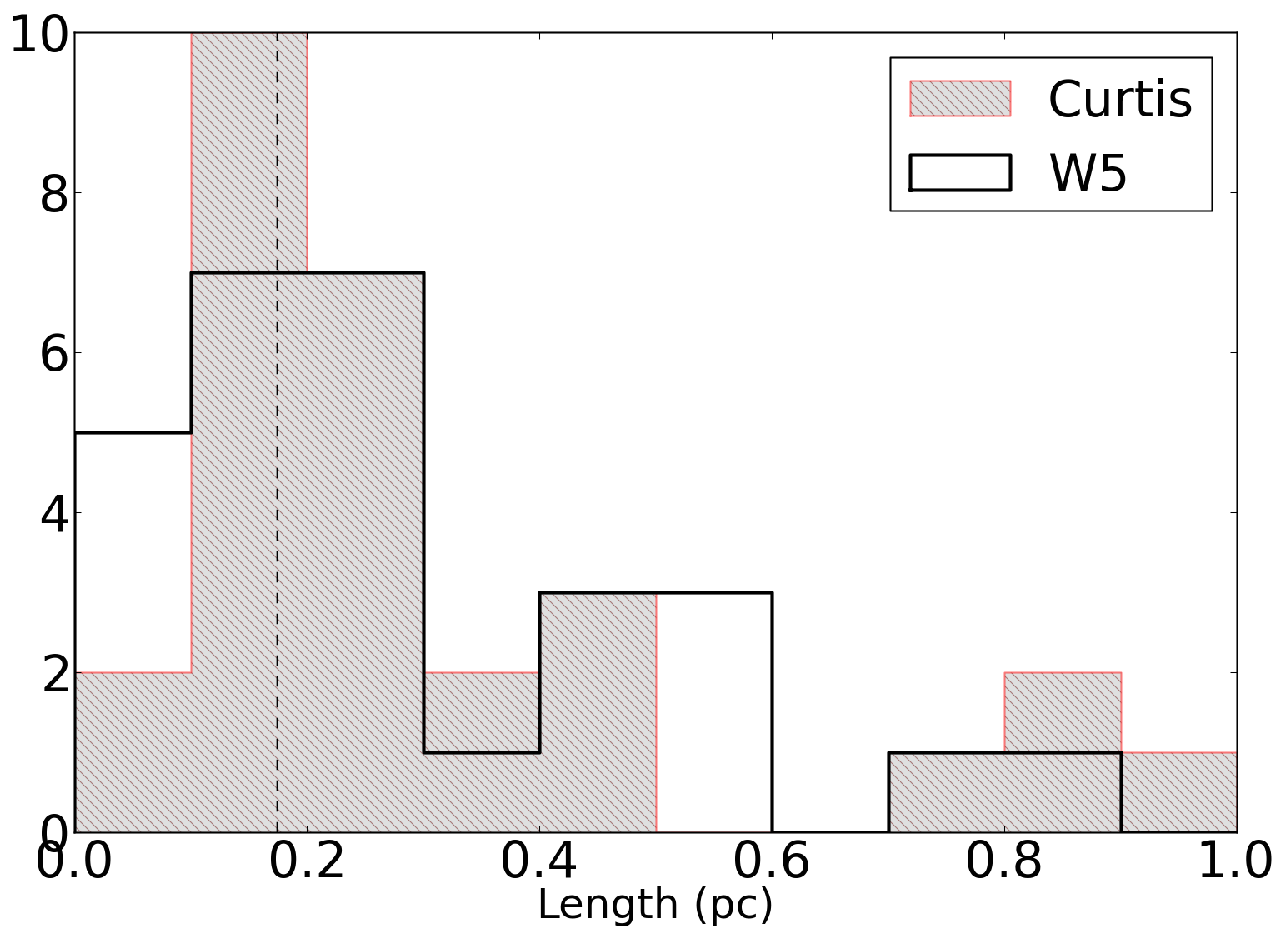}
{Histogram of the measured outflow lobe separations.  The grey hatched region shows
\citet{curtis2010} values.  The vertical dashed line represents the spatial
resolution of our survey.  The two distributions are similar.}
{fig:lengthhist}{1.0}

On average, the \citet{curtis2010} outflow velocities are similar to ours
(Figure \ref{fig:widthhist}).   We detect lower velocity outflows because we do
not set a strict lower velocity limit criterion.  We do not detect the highest
velocity outflows most likely because of our poorer sensitivity to the faint
high-velocity tips of outflows, although it is also possible that no
high-velocity ($v>20$ \kms) flows exist in the W5 region.  Note that the
histogram compares quantities that are not directly equivalent: the outflows in
\citet{curtis2010} and our own data are measured out to the point at which the
outflow signal is lost, while the `region' velocities are full-width half-max
(FWHM) velocities.  

Finally, we use the detectability of outflows in Perseus to inform our
expectations in W5.  Since it appears that we can detect outflows from low-mass
protostars with sub-stellar to $\sim30L_\odot$ luminosities
at the distance of W5 and these objects should be the most numerous in a
standard initial mass function, the distribution of physical properties in W5
outflows should be similar to those in Perseus.  However, because W5 is a
somewhat more massive cloud ($M_{W5}\approx 5 M_{Perseus}$ \footnote{$M_{W5}$
is estimated from \thirteenco.  We also estimate the total molecular mass in W5 using
the X-factor and acquire $M_{W5}=5.0\ee{4}$ \msun, in agreement with
\citet{karr:triggered:2003}, who estimated a molecular mass of 4.4\ee{4} from
\twelveco\ using the same X-factor.  \citet{koenig:clustered:2008} estimated a
total gas mass of 6.5\ee{4} from a 2MASS extinction map.   The total molecular mass
in Perseus is $M_{Perseus} \sim 10^4$ \citep{bally-perseus2008}}), we might expect the
high-end of the distribution to extend to higher values of outflow mass,
momentum, and energy.  Since we will likely see clustered outflows confused
into a smaller number of distinct lobes, we expect a bias towards higher values
of the derived quantities but a lower detection rate.

\subsubsection{Velocity, Column Density, and Mass Measurements}
\label{sec:measurements}
Throughout this section, we assume that the CO
lines are optically thin and thermally excited.   The measured properties
are presented in Table \ref{tab:outflows}.  These assumptions are
likely to be invalid, so we also discuss the consequences of applying `typical'
optical depth corrections to the derived quantities.   Because we do
not measure optical depths and the optical depth correction for CO 3-2 is less
well quantified than for CO 1-0 \citep{curtis2010,Cabrit1990}\footnote{In
\citet{curtis2010}, this correction factor ranged from 1.8 to 14.3;
\citet{arce2010} did not enumerate the optical depth correction they
used but it is typically around 7 \citep{Cabrit1990}.  }, we only present the
uncorrected measurements in Table \ref{tab:outflowsderived}.

The outflow velocity ranges were measured by examining
both RA-velocity and Dec-velocity diagrams interactively using the STARLINK
GAIA data cube viewing tool.  The velocity limits are set to include
all outflow emission that is distinguishable from the cloud (i.e. the velocity
at which outflow lobes dominate over the gaussian wing of the cloud
emission) down to zero emission.  An outflow size \citep[or lobe size,
following ][]{curtis2010} was determined by integrating over the blue and red
velocity ranges and creating an elliptical aperture to include both peaks; the
position and size therefore have approximately beam-sized ($\approx18\arcsec$)
accuracy.  The integrated outflow maps are shown as red and blue contours in
Figure \ref{fig:pv2b}.  The velocity center was computed by fitting a
gaussian to the FCRAO \thirteenco\ spectrum averaged over the elliptical
aperture.

\Figure{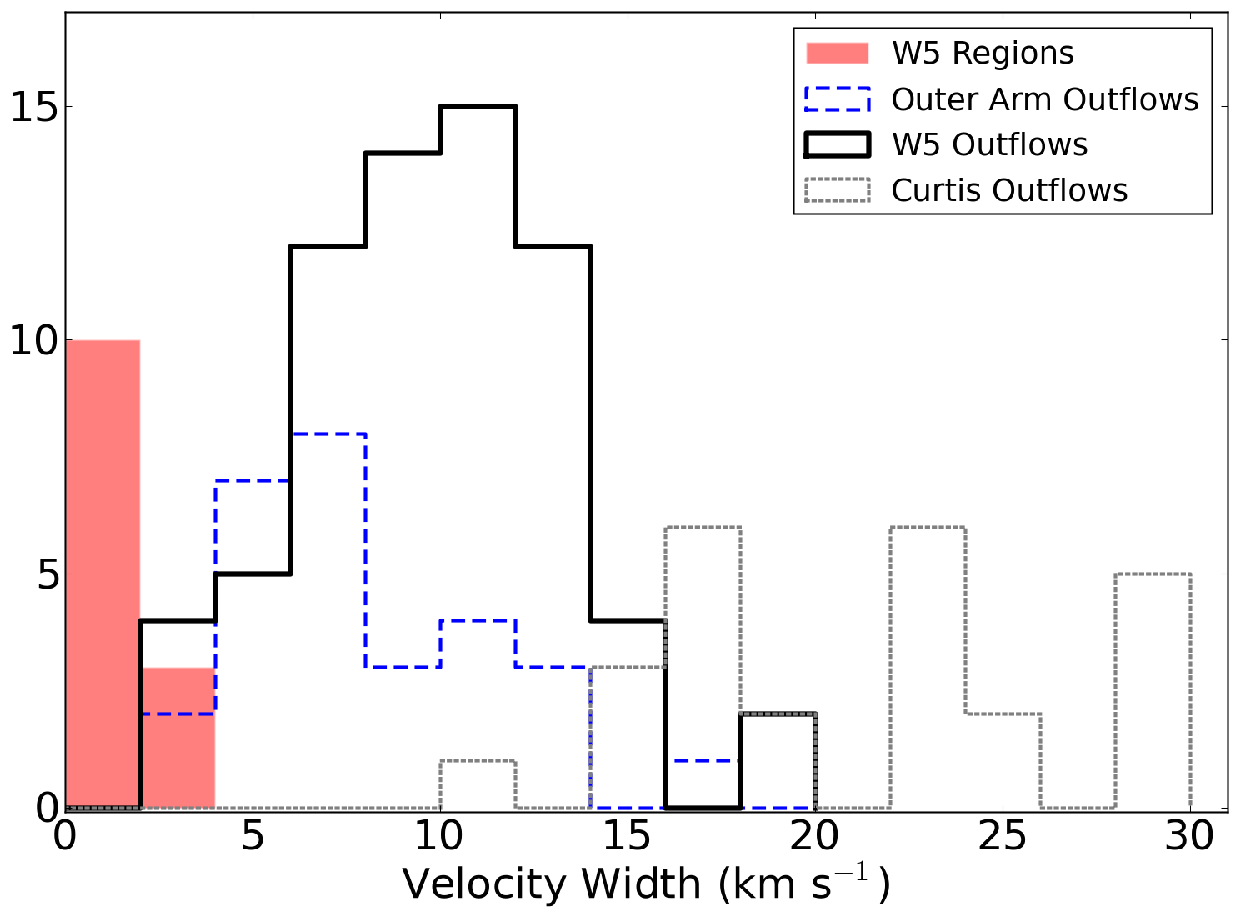}
{Histogram of the outflow line widths. {\it Black lines}: histogram of the measured
outflow widths (half-width zero-intensity, measured from the fitted central
velocity of the cloud to the highest velocity with non-zero emission).  {\it
Blue dashed lines}: outflow half-width zero-intensity (HWZI) for the outer arm (non-W5) sample.
{\it Solid red shaded}: The measured widths (HWHM) of the sub-regions as
tabulated in Table \ref{tab:regionspectra}.   
{\it Gray dotted}: Outflow $v_{max}$ (HWZI) values for Perseus
from \citet{curtis2010}. }
{fig:widthhist}{1.0}

The column density is estimated from \twelveco\ J=3-2  assuming local thermal
equilibrium (LTE) and optically thin emission using the equation 
$ N(\hh) =
5.3\ee{18}\eta_{mb}^{-1} \int T_A^*(v) dv $ for $T_{ex}=20$ K. 
The derivation is given in the Appendix.
The column density in the lobes is likely to be dominated by low-velocity gas
and therefore our dominant uncertainty may be missing low-velocity emission
rather than poor assumptions about the optical depth.

The scalar momentum and energy were computed from
\begin{equation}
      p = M \frac{\sum T_A^*(v) (v-v_{c}) \Delta v}{ \sum T_A^*(v) \Delta v}
\end{equation}
\begin{equation}
      E = \frac{M}{2} \frac{\sum T_A^*(v) (v-v_{c})^2 \Delta v}{ \sum T_A^*(v) \Delta v}
\end{equation}
where $v_c$ is the \thirteenco\ 1-0 centroid velocity.  The same 
assumptions used in determining column density are applied here.

We
estimate an outflow lifetime by taking half the distance between the red and
blue outflow centroids divided by the maximum measured velocity difference
($\Delta v_{max} = (v_{max,red}-v_{max,blue})/2$), $\tau_{flow} = L_{flow} / ( 2 \Delta
v_{max})$, where $L_{flow}$ refers to the length of the flow.  This method
assumes that the outflow inclination is 45\degree; if it is more parallel to
the plane of the sky, we overestimate the age, and vice-versa.  The momentum
flux is then $\dot{P} = p / \tau$.  Similarly, we compute a mass loss rate by
dividing the total outflow mass by the dynamical age, which yields what is
likely a lower limit on the mass loss rate (if the lifetime is underestimated,
the mass loss rate is overestimated, but the outflow mass is always a lower
limit because of optical depth and confusion effects).    

The dynamical ages are highly suspect since the red and blue lobes are often
unresolved or barely resolved, and diffuse emission averaged with the lobe
emission can shift the centroid position.  Additionally, it is not clear what
portion of the outflow corresponds to the centroid: the bow shock or the jet could both potentially
dominate the outflow emission.  \citet{curtis2010} discuss the many ways in
which the dynamical age can be in error.  
Our mass loss rates are similar to those in Perseus \emph{without} correcting our 
measurements for optical depth, while our outflow masses are an order of magnitude lower.
It therefore appears that our dynamical age estimates must be too low, since we have no 
reason to expect protostars in W5 to be undergoing mass loss at a greater rate than those in
Perseus.
However, given more reliable
dynamical age estimates from higher resolution observations of shock tracers,
the mass loss rates could be corrected and compared to other star-forming
regions.

Because the emission was assumed to be optically thin, the mass, column,
energy, and momentum measurements we present are strictly lower limits.  While
some authors have computed correction factors to \twelveco\ 1-0 optical depths
\citep[e.g.][]{Cabrit1990},  the corrections are different for the 3-2
transition \citep[1.8 to 14.3,][]{curtis2010}.  Additionally, CO 3-2 may
require a correction for sub-thermal excitation because of its higher critical
density (the CO 3-2 critical density is 27 times higher than CO 1-0; see
Appendix \ref{appendix:dipole} for modeling of this effect).

Additionally, most of the outflow mass is at the lowest distinguishable
velocities in typical outflows \citep[e.g.][]{arce2010}.  It is therefore
plausible that in the more turbulent W5 region, a greater fraction of the
outflow mass is blended (velocity confused) with the cloud and therefore not
included in mass, momentum, and energy measurements.  This omission could be
greater than the underestimate due to poor opacity assumptions.

The total mass of the W5 outflows is $M_{tot}\approx1.5 \msun$,
substantially lower, even with an optical depth correction of $10\times$, than
the 163 \msun\ reported in Perseus \citep{arce2010}.  \citet{arce2010} also include
a correction factor of 2.5 to account for higher temperatures in outflows and a
factor of 2 to account for emission blended with the cloud.  The temperature
correction is inappropriate for CO 3-2 (see Appendix \ref{appendix:dipole},
Figure \ref{fig:approx}), but the resulting total outflow mass in W5 with an
optical depth correction and a factor of 2 confusion correction is about 30
\msun.  In order to make our measurements consistent with a mass of 160 \msun\ , a density
upper limit in the outflowing gas of $n(\hh) < 10^{3.5} \percc$ is required,
since a lower gas density results in greater mass for a given intensity (see
Appendix \ref{appendix:dipole}, Figure \ref{fig:coradex}).  However, we
expect the total outflow mass in W5 to be greater than in Perseus because 
of the greater cloud mass, implying that the density in the flows must be even
lower, or additional corrections are needed.

The total outflow momentum is $p_{tot}\approx10.9 \msun$ \kms, versus a quoted
517 \msun \kms\ in Perseus \citep{arce2010}.  \citet{arce2010} included
inclination and dissociative shock corrections for the momentum measurements
on top of the correction factors already applied to the mass.  If these
corrections are removed from the Perseus momentum total (except for optical depth,
which is variable in their data and therefore cannot be removed), the uncorrected outflow
momentum in Perseus would be about 74 \msun \kms. The W5 outflow momentum, if
corrected with a `typical' optical depth in the range 7-14, would match or exceed
this value.  If an additional CO 3-2 excitation correction (in the range 1-20)
is applied, the W5 outflow momentum would significantly exceed that in Perseus.

Assuming a turbulent line width $\Delta v \sim 3$ \kms\ (approximately the
smallest FWHM line-width observed), the total turbulent momentum in the ambient
cloud is $p = M_{tot} \Delta v = 1.3\ee{5} \msun$ \kms, which is $\sim10^5$
times the measured outflow momentum - the outflows detected in our
survey cannot be the sole source of the observed turbulent line widths, even if
corrected for optical depth and missing mass.  

Table \ref{tab:regionspectra} presents the turbulent momentum for each
sub-region computed by multiplying the measured velocity width by the
integrated \thirteenco\ mass.  Even if the outflow measurements are 
orders of magnitude low because of optical depth, cloud blending, sub-thermal
excitation, and other missing-mass considerations, outflows contribute
negligibly to the total momentum of high velocity gas in W5.  This result is
unsurprising, as there are many other likely sources of energy in the region
such as stellar wind bubbles and shock fronts between the ionized and molecular
gas.  Additionally, in regions unaffected by feedback from the HII region (e.g.
W5NW), cloud-cloud collisions are a possible source of energy.

Figure \ref{fig:outflowhist} displays the distribution of measured properties
and compares them to those derived in the COMPLETE \citep{arce2010} and
\citet{curtis2010} HARP CO 3-2 surveys of Perseus.  Our derived masses are
substantially lower than those in \citet{arce2010} even if corrected for
optical depth, but our momenta are similar to the CPOC (COMPLETE Perseus
Outflow Candidate) sample and our energies are higher, indicating a bias
towards detecting mass at high velocities.  The bias is more heavily towards
high velocities than the CO 1-0 used in \citet{arce2010}.  The discrepancy
between our values and those of \citet{arce2010} and \citet{curtis2010} can be
partly accounted for by the optical depth correction applied in those works:
\thirteenco\ was used to correct for opacity at low velocities, where most of
the outflow mass is expected.  Those works may also have been less affected by
blending because of the smaller cloud line widths in Perseus.

\FigureFour{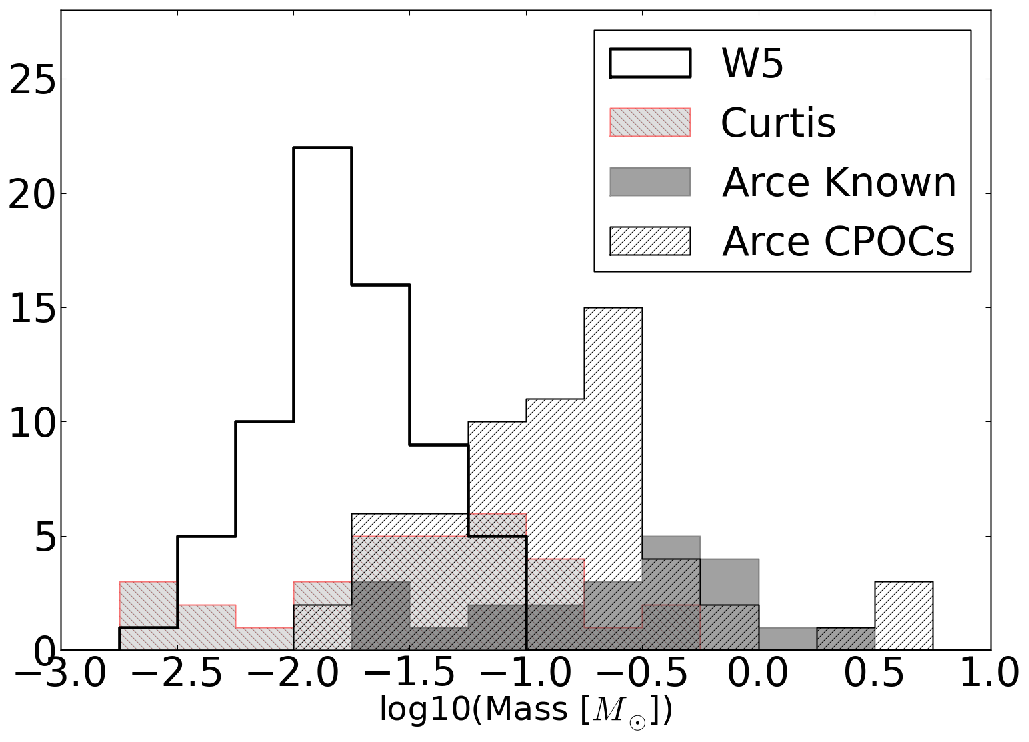}{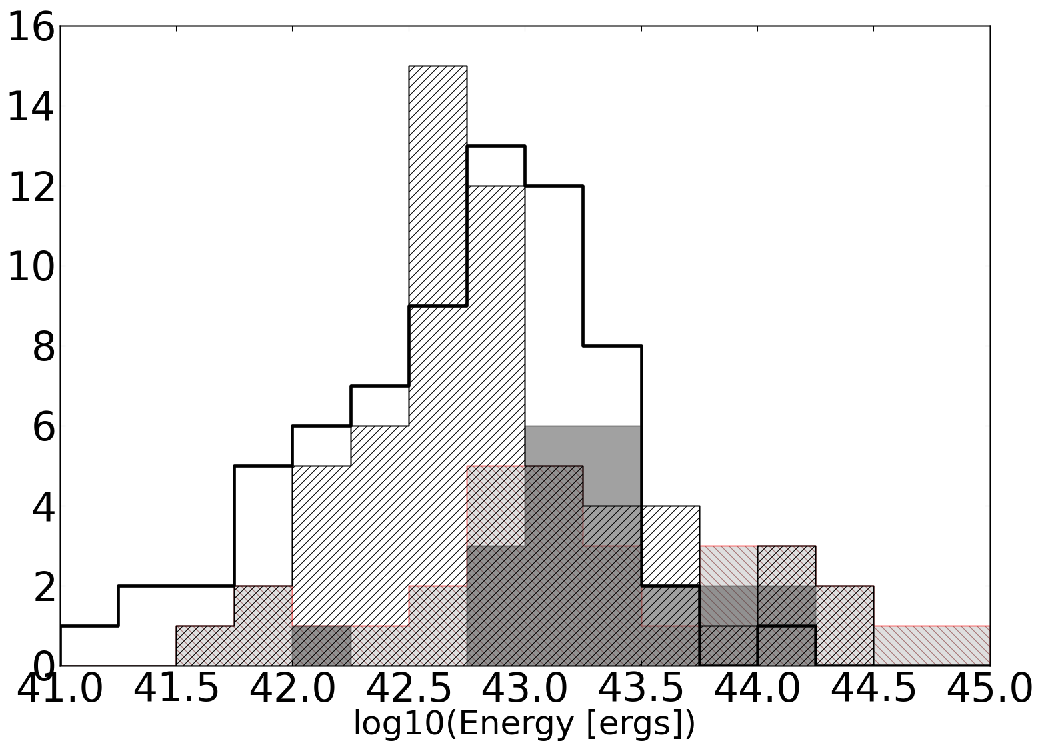}{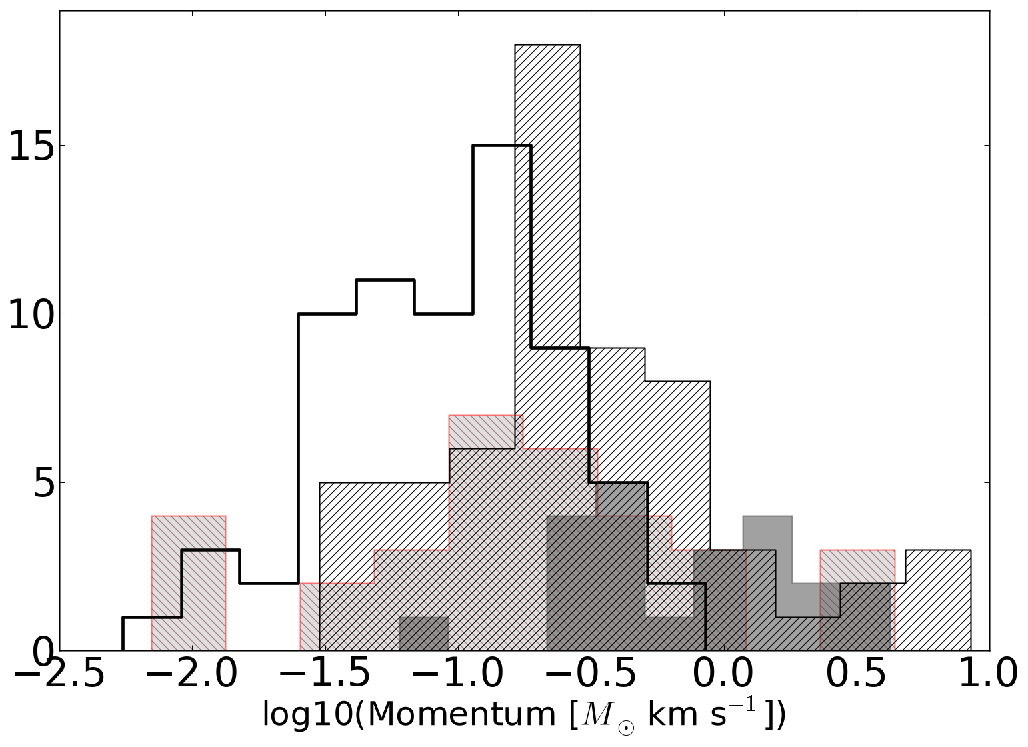}{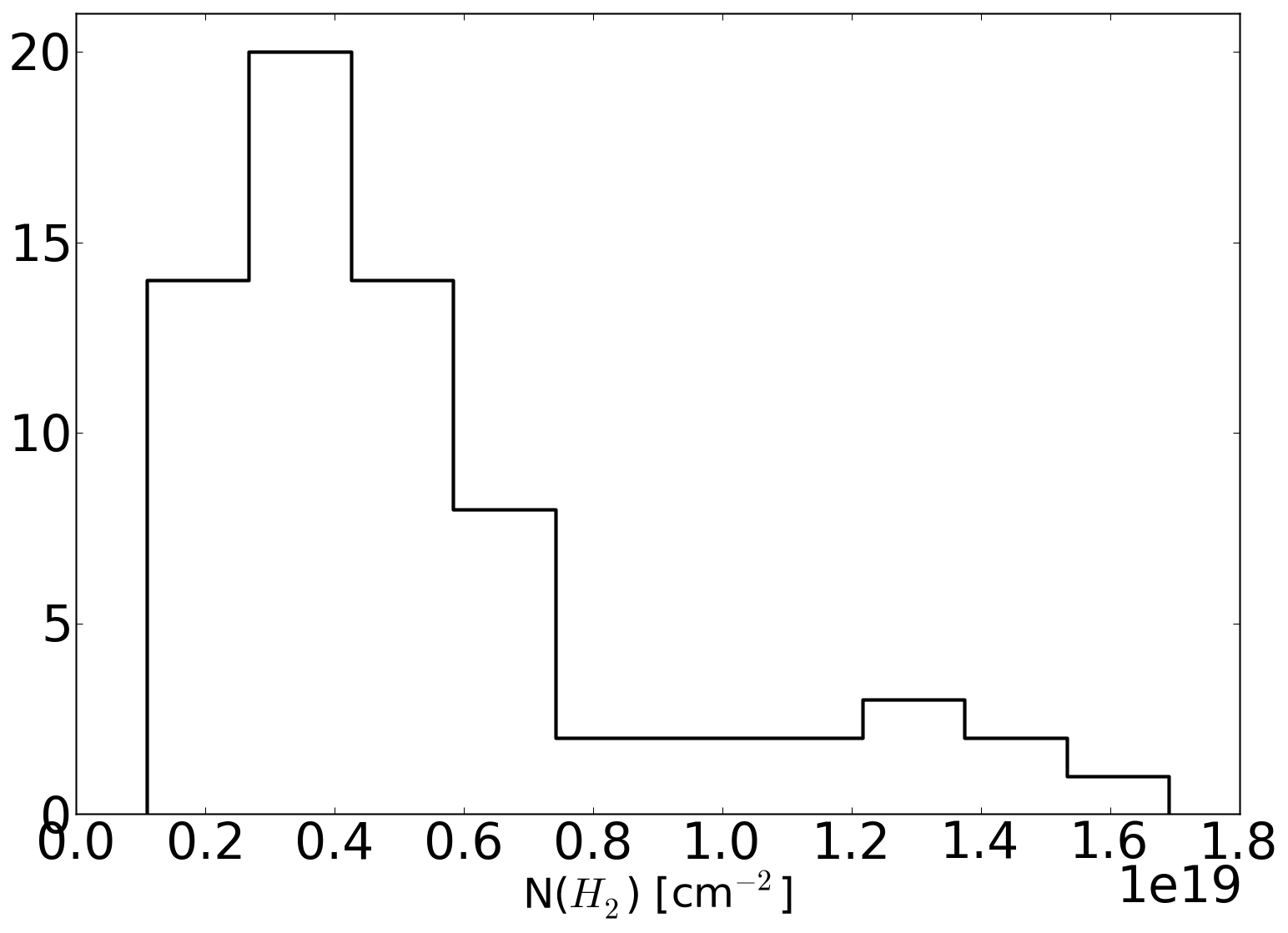}
{Histograms of outflow physical properties.  
The solid unfilled lines are the W5 outflows (this paper), the forward-slash
hashed lines show \citet{arce2010} CPOCs , the dark gray
shaded region shows \citet{arce2010} values for known outflows in Perseus, and
the light gray, backslash-hashed regions show \citet{curtis2010} CO 3-2 outflow
properties.  The outflow masses measured in Perseus are systematically higher
partly because both surveys corrected for line optical depth using \thirteenco.
The medians of the distributions are 0.017, 0.044, 0.33, and 0.14 \msun\ for
W5, Curtis, Arce Known, and Arce CPOCs respectively, which implies that an
optical depth and excitation correction factor of 2.5-20 would be required to
make the distributions agree (although W5, being a more massive region, might
be expected to have more massive and powerful outflows).  It is likely that CO
3-2 is sub-thermally excited in outflows, and CO outflows may be destroyed by
UV radiation in the W5 complex while they easily survive in the lower-mass
Perseus region, which are other factors that could push the W5 mass
distribution lower.
}
{fig:outflowhist}

The momentum flux and mass loss rate are compared to the values derived in
Perseus by \citet{hatchell2007} and \citet{curtis2010} in Figures
\ref{fig:outflowPflux} and \ref{fig:outflowmdot}.  Both of our values are
computed using the dynamical timescale $\tau_d$ measured from outflow lobe
separation, while the \citet{hatchell2007} values are derived using a more
direct momentum-flux measurement in which the momentum flux contribution 
of each pixel in the resolved outflow map is considered.  
The derived
momentum fluxes (Figure \ref{fig:outflowPflux}) are approximately consistent
with the \citet{curtis2010} Perseus momentum fluxes; \citet{curtis2010} measure
momentum fluxes in a range $1\ee{-6}<\dot{P}<7\ee{-4}$ \msun \kms \peryr,
higher than our measured $6\ee{-7}<\dot{P}<1\ee{-4}$ \msun \kms \peryr\ by
approximately the opacity correction they applied.  As seen in Figure
\ref{fig:outflowPflux}, the \citet{hatchell2007} momentum flux measurements in
Perseus cover a much lower range $6\ee{-8}<\dot{P}<2\ee{-5}$ \msun \kms \peryr\
and are not consistent with our measurements.  This disagreement is most likely
because of the difference in method.  The W5 outflows exhibit
substantially higher mass-loss rates and momentum fluxes if we assume a factor
of 10 opacity correction, as expected from our bias toward higher-velocity,
higher-mass flows.

\Figure{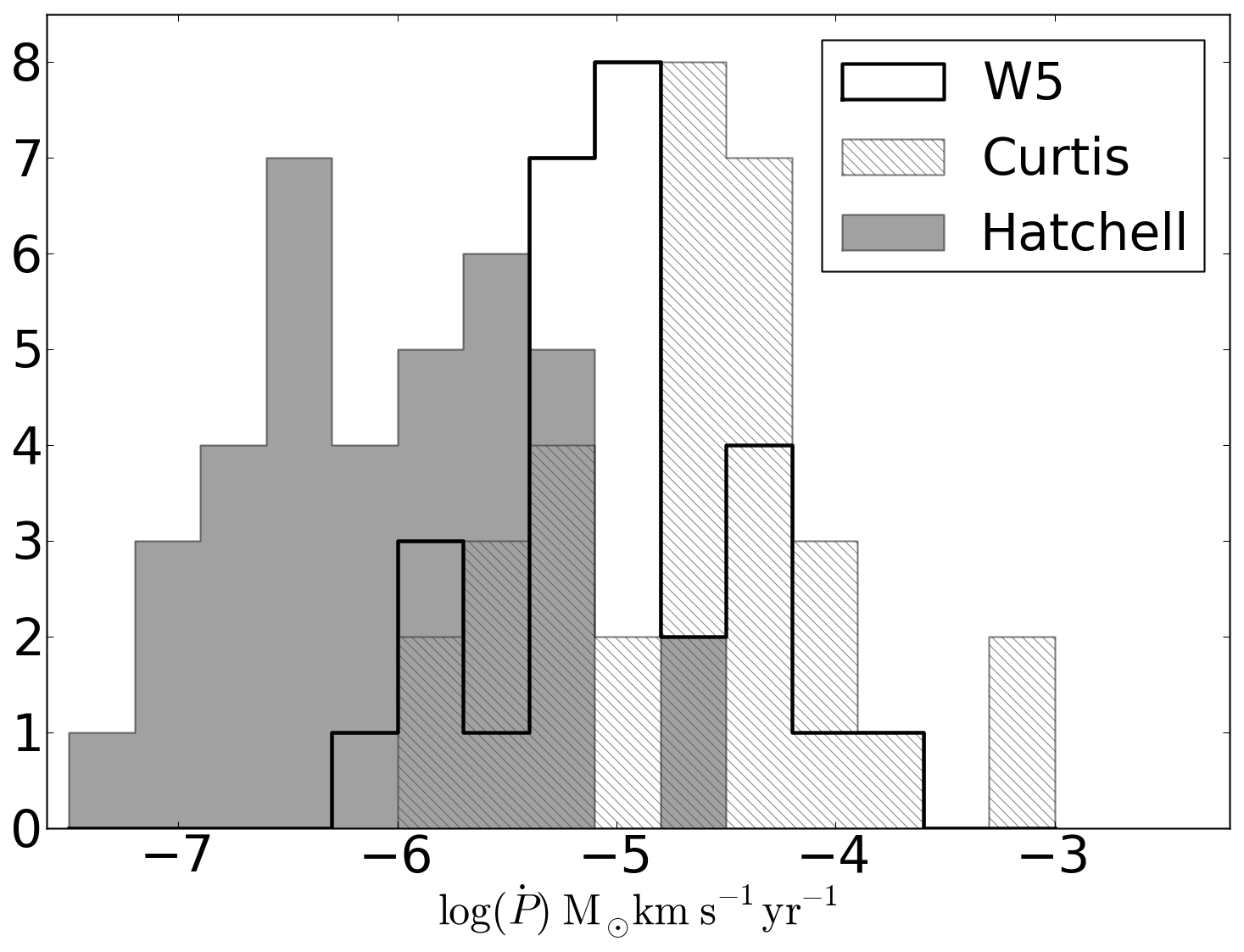}
{Histogram of the measured outflow momentum fluxes.  The black thick line shows
our data, the grey shaded region shows the \citet{hatchell2007} data, and the
hatched region shows \citet{curtis2010} values.  Our measurements peak squarely
between the two Perseus JCMT CO 3-2 data sets, although the \citet{curtis2010}
results include an opacity correction that our data do not, suggesting that our
results are likely consistent with \citet{curtis2010} but inconsistent with the
\citet{hatchell2007} direct measurement method.}
{fig:outflowPflux}{1.0}

\Figure{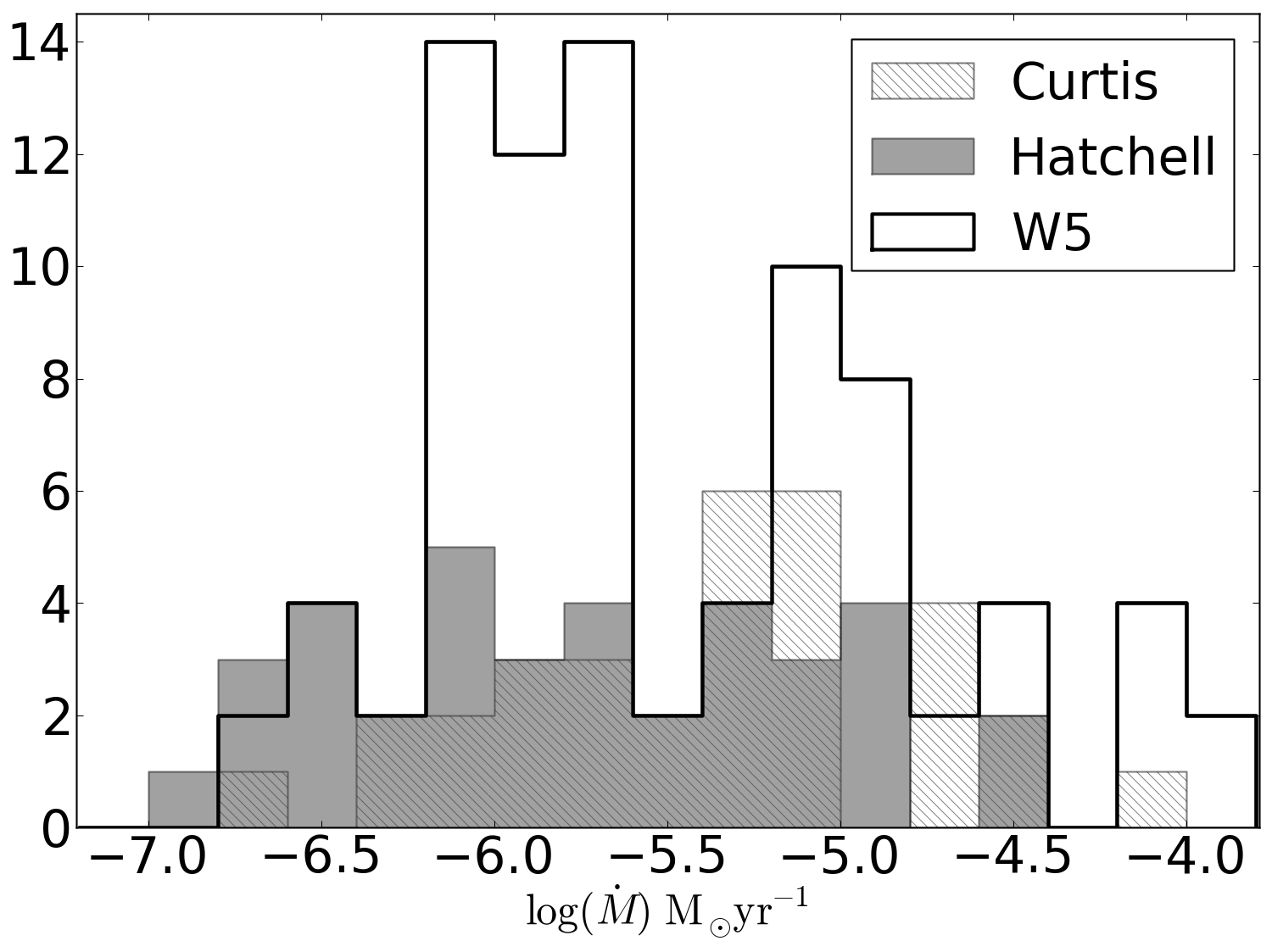}
{Histogram of the measured mass loss rate.  The black thick line shows our
data, while the grey shaded region shows the \citet{hatchell2007} data, which
is simply computed by $\dot{M} = \dot{P} \times 10 / 5$ \kms, where the factor
of 10 is a correction for opacity.  Our mass loss rates are very comparable to
those of \citet{hatchell2007}, but different methods were used so the
comparison may not be physically meaningful.  \citet{curtis2010} (hatched) used
a dynamical time method similar to our own and also derived similar mass loss
rates, although their mass measurements have been opacity-corrected using the
\thirteenco\ 3-2 line.  Because our mass loss rates agree reasonably with
Perseus, but our outflow mass measurements are an order of magnitude low, we
believe our dynamical age estimates to be too small.
}
{fig:outflowmdot}{1.0}

\subsection{Structure of the W5 molecular clouds: A thin sheet?}
The W5 complex extends $\sim 1.6\degree \times 0.7 \degree$ within 20\degree\
of parallel with the galactic plane.  At the assumed distance of 2 kpc, it has
a projected length of $\sim60$ pc (Figure \ref{fig:outflows_on_co32}).
In the 8 \um\ band (Figure \ref{fig:color_overview}), the region appears to
consist of two blown-out bubbles with $\sim 10-15$ pc radii centered on
$\ell=138.1, b=1.4$ and $\ell=137.5,b=0.9$.  While the bubbles are filled in
with low-level far-infrared emission, there is no CO detected down to a
$3-\sigma$ limit of 3.0 K \kms\ (\twelveco\ 1-0), 2.4 K \kms\ (\twelveco\ 3-2,
excepting a few isolated clumps), and 1.5 K \kms\ (\thirteenco\ 1-0).  Using the 
X-factor (the CO-to-\hh\ conversion factor) for \twelveco\, $N(\hh) =
3.6\ee{20} \persc / (\mathrm{K}~\kms)$, we derive an upper limit $N(\hh) <
1.1\ee{21}\persc$, or $A_V \lesssim 0.6$.  Individual `wisps' and `clumps' of
CO can sometimes be seen, particularly towards the cloud edges, but in general
the bubbles are absent of CO gas.

Given such low column limits,  the W5 cloud must be much smaller along the line
of sight than its $\sim50$ pc size projected on the sky.  Alternately, along
the line-of-sight, the columns of molecular gas are too low for CO to
self-shield, and it is therefore destroyed by the UV radiation of W5's O-stars.
In either case, there is a significant excess of molecular gas in the plane of
the sky compared to the line of sight, which makes W5 an excellent location to
perform unobscured observations of the star formation process.  The implied
thin geometry of the W5 molecular cloud may therefore be similar to the bubbles
observed by \citet{Beaumont2009}, but on a larger scale.

There is also morphological evidence supporting the face-on
hypothesis.  In the AFGL 4029 region (Section \ref{sec:afgl4029}) and all along
the south of W5, there are ridges with many individual cometary `heads'
pointing towards the O-stars that are unconfused along the line of sight.  This
sort of separation would not be expected if we were looking through the clouds
towards the O-stars.  W5W, however, presents a counterexample in which there
are two clouds along the line of sight that may well be masking a more complex
geometry.

\section{Sub-regions}
\label{sec:subregions}

Individual regions were selected from the mosaic for comparison.  All regions
with multiple outflows and indicators of star formation activity were named and
included as regions for analysis.  Additionally, three ``inactive'' regions were
selected based on the presence of \thirteenco\ emission but the lack of
outflows in the \twelveco\ 3-2 data.  Finally, two regions devoid of CO emission
were selected as a baseline comparison and to place upper limits on the molecular
gas content of the east and west `bubbles'.  The regions are identified on the
integrated \thirteenco\ image in Figure \ref{fig:regionboxes_on_CO}. 

Average spectra were taken of each ``region'' within the indicated box.
Gaussians were fit to the spectrum to determine line-widths and centers (Figure
\ref{fig:regionspectra}, Table \ref{tab:regionspectra}).  Gaussian fits were
necessary because in many locations there are at least two velocity components,
so the second moment (the ``intensity-weighted dispersion'') is a poor
estimator of line width.  Widths ranged from $v_{FWHM} = 2.3$ to 6.2 \kms\
(Figure \ref{fig:widthhist}).

\Table{cccccccccc}{Gaussian fit parameters of sub-regions}
{{Region} & {Velocity 1} & {Width 1} & {Amplitude 1} & {Velocity 2} & {Width 2} & {Amplitude 2} &  &  & \\
 & {(\kms)} & {(FWHM, \kms)} & {(K)} & {(\kms)} & {(FWHM, \kms)} & {(K)} &  & \\}
{tab:regionspectra}
{
S201 & -38.04 & 3.149 & 2.35 & - & - & -\\
AFGL4029 & -38.91 & 3.3605 & 1.48 & - & - & -\\
LWCas & -38.83 & 3.478 & 2.33 & - & - & -\\
W5W & -41.37 & 3.8775 & 3.07 & -36.16 & 3.8305 & 1.90\\
W5NW & -36.37 & 3.854 & 1.6 & - & - & -\\
W5NWpc & -36.37 & 3.713 & 1.19 & -41.81 & 4.3475 & 0.47\\
W5SW & -42.78 & 4.136 & 0.6 & -36.34 & 4.183 & 0.22\\
W5S & -40.15 & 2.914 & 0.34 & -35.76 & 2.2795 & 0.40\\
Inactive1 & -42.91 & 2.6555 & 0.75 & -39.38 & 4.2065 & 0.42\\
Inactive2 & -38.94 & 3.7365 & 1.2 & - & - & -\\
empty & -37.81 & 5.217 & 0.04 & - & - & -\\
\hline \\
\thirteenco\ fits &&&&&&& \thirteenco      & \thirteenco          & \thirteenco       \\
                   &&&&&&&              mass &              momentum &              energy\\
                   &&&&&&&(\msun)           &(\msun \kms)         &(ergs)              \\
\hline \\
S201 & -37.97 & 2.5615 & 0.56 & - & - & - & 1300 & 3500 & 8.9\ee{46}\\
AFGL4029 & -38.66 & 2.35 & 0.35 & - & - & - & 2600 & 6100 & 1.4\ee{47}\\
LWCas & -38.75 & 2.679 & 0.51 & - & - & - & 3700 & 10000 & 2.7\ee{47}\\
W5W & -41.23 & 2.773 & 1.09 & -36.51 & 3.5485 & 0.47 & 4500 & 13000 & 3.5\ee{47}\\
W5NW & -36.1 & 3.431 & 0.7 & - & - & - & 5300 & 18000 & 6.3\ee{47}\\
W5NWpc & -36.18 & 3.3135 & 0.42 & -41.44 & 3.619 & 0.14 & 15000 & 50000 & 1.6\ee{48}\\
W5SW & -42.6 & 3.807 & 0.1 & -36.15 & 4.2535 & 0.05 & 790 & 3000 & 1.1\ee{47}\\
W5S & -39.9 & 2.444 & 0.07 & -35.48 & 2.209 & 0.08 & 320 & 790 & 1.9\ee{46}\\
Inactive1 & -42.58 & 2.5145 & 0.1 & -38.97 & 2.82 & 0.07 & 1400 & 3500 & 8.7\ee{46}\\
Inactive2 & -38.82 & 3.196 & 0.37 & - & - & - & 3100 & 9900 & 3.2\ee{47}\\
empty & -38.44 & 4.7705 & 0.02 & - & - & - & 340 & 1600 & 7.8\ee{46}\\
}{
}

\Figure{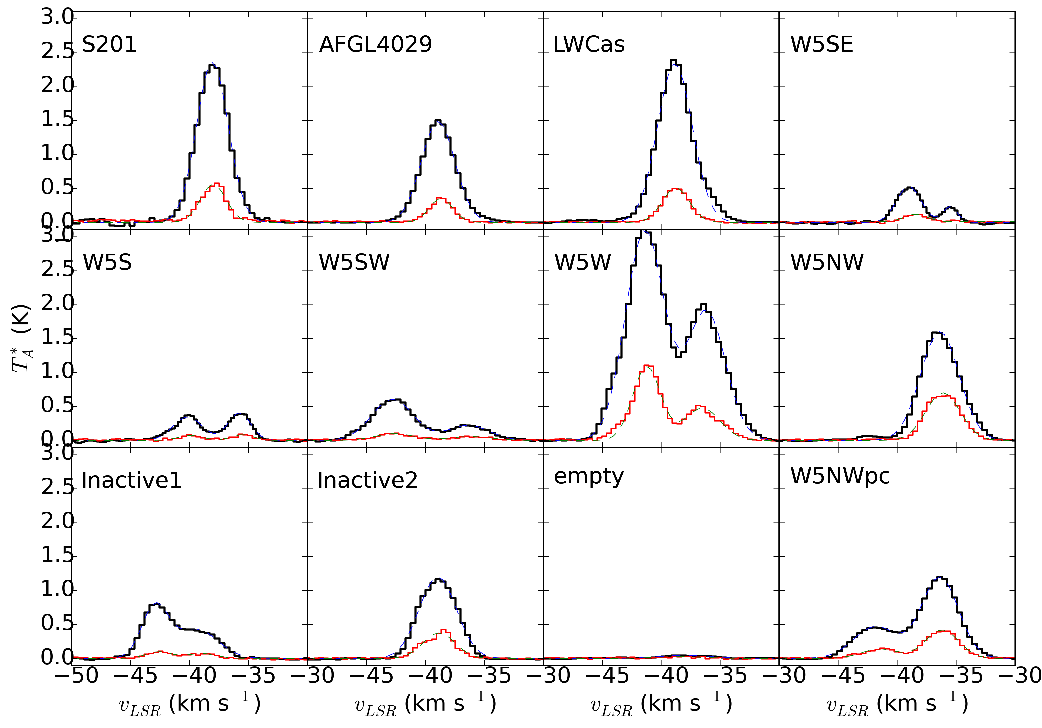}
{Spatially averaged spectra of the individual regions analyzed.  \twelveco\ 3-2
is shown by thick black lines and \thirteenco\ 1-0 is shown by thin red lines.
Gaussian fits are overplotted in blue and green dashed lines, respectively.
The fit properties are given in Table
\ref{tab:regionspectra}.}{fig:regionspectra}{3.0}

\subsection{Sh 2-201}
Sh 2-201 is an HII region and is part of the same molecular cloud as the
bright-rimmed clouds in W5E, but it does not share a cometary shape with these
clouds (Figure \ref{fig:S201}).  Instead, it is internally heated and has its
own ionizing source \citep{Felli1987}.  The AFGL 4029 cloud edge is at a projected
distance of $\sim7$ pc from the nearest exposed O-star, and the closest
illuminated point in the Spitzer 8 and 24 \um\ maps is at a projected distance
of $\sim 5$ pc.  The star forming process must therefore have begun before
radiation driven shocks from the W5 O-stars could have impacted the cloud.  

\Figure{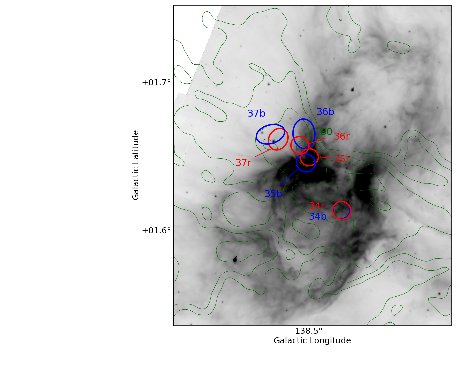} 
{Small scale map of the Sh 2-201 region plotted with CO 3-2 contours integrated
from -60 to -20 \kms\ at levels 3,7.2,17.3,41.6, and 100 K \kms.  
The IRAC 8 \um\ image is displayed in inverted log scale from \lowirac\
to \highirac\ MJy \persr. Contours of the CO 3-2 cube integrated from
-60 to -20 \kms\ are overlaid at logarithmically spaced levels from 3 to 100 K
\kms\ (3.0,7.2,17.3,41.6,100; $\sigma\approx0.7$ K \kms).  The
ellipses represent the individual outflow lobe apertures mentioned in Section
\ref{sec:measurements}.
}{fig:S201}{1.0}

\subsection{AFGL 4029}
\label{sec:afgl4029}
AFGL 4029 is a young cluster embedded in a cometary cloud (Figure
\ref{fig:afgl4029}).  There is one clear bipolar outflow and 6 single-lobed
flows that cannot be unambiguously associated with an opposite direction
counterpart.  The cluster is mostly unresolved in the data presented here and
is clearly the most active CO clump in W5.  It contains a cluster of at least 30 B-stars
\citep{Deharveng1997}.  The outflows from this region have a full width $\Delta
v \approx 30$ \kms, which is entirely inconsistent with a radiation-driven
inflow or outflow since it is greater than the sound speed in the ionized
medium.

The northeast cometary cloud is strongly affected by
the W5 HII region.  It has an outflow in the head of the cloud (Figure
\ref{fig:necomet}), and the cloud shows a velocity gradient with distance from
the HII region.  The polarity of the gradient suggests that the cometary cloud
must be on the far side of the ionizing O-star along the line of sight assuming
that the HII region pressure is responsible for accelerating the cloud edge.

\Figure{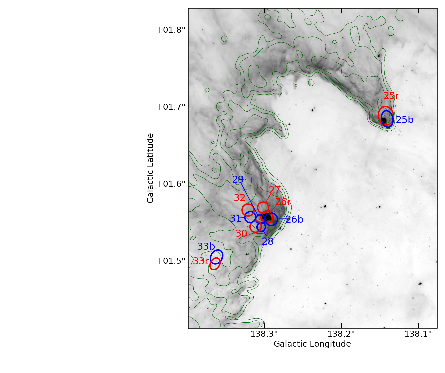}
{Small scale map of the AFGL 4029 region plotted with CO 3-2 contours integrated
from -60 to -20 \kms\ at levels 3,7.2,17.3,41.6, and 100 K \kms.  
The IRAC 8 \um\ image is displayed in inverted log scale from \lowirac\ to
\highirac\ MJy \persr. Contours of the CO 3-2 cube integrated from -60 to -20
\kms\ are overlaid at logarithmically spaced levels from 3 to 100 K \kms\
(3.0,7.2,17.3,41.6,100; $\sigma\approx0.7$ K \kms).  Outflows 26-32 are ejected
from a forming dense cluster.  A diagram displaying the kinematics of the
northern cometary cloud is shown in Figure \ref{fig:necomet}. }
{fig:afgl4029}{1.0}

\Figure{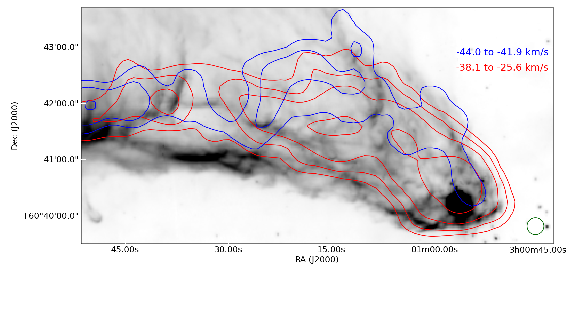}
{The northeast cometary cloud.  Contours are shown at 0.5,1,2, and 5 K \kms\
integrated over the ranges -44.0 to -41.9 \kms\ (blue) and -38.1 to -35.6 \kms\
(red).  There is a velocity gradient across the tail, suggesting that the front
edge is being pushed away along the line of sight.}
{fig:necomet}{1.0}

\subsection{W5 Ridge}
\label{sec:w5ridge}
The W5 complex consists of two HII region bubbles separated by a ridge of
molecular gas (Figure \ref{fig:lwcas}).  This ridge contains the LW Cas optical
nebula, a reflection nebula around the variable star LW Cas, on its east side
and an X-shaped nebula on the west.  The east portion of LW Cas Nebula is
bright in both the continuum and CO J=3-2 but lacks outflows (see Figure
\ref{fig:lwcas}).  The east portion also has the highest average peak antenna
temperature, suggesting that the gas temperature in this region is
substantially higher than in the majority of the W5 complex (higher spatial
densities could also increase the observed $T_A$, but the presence of nearby heating
sources make a higher temperature more plausible).  It is possible
that high gas temperatures are suppressing star formation in the cloud.
Alternately, the radiation that is heating the gas may destroy any outflowing
CO, which is more likely assuming the two Class I objects identified in this
region by \citet{koenig:clustered:2008} are genuine protostars.

\Figure{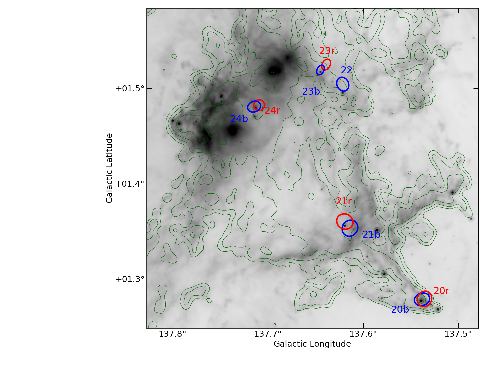} 
{Small scale map of the LW Cas nebula plotted with CO 3-2 contours integrated
from -60 to -20 \kms\ at levels 3,7.2,17.3,41.6, and 100 K \kms.  The feature containing outflows 20 and
21 is the X-shaped ridge referenced in Section \ref{sec:w5ridge}.  This
sub-region is notable for having very few outflows associated with the most
significant patches of CO emission.   The gas
around it is heated on the left side by the O7V star HD 18326 ($D_{proj}=8.5$
pc), suggesting that this gas could be substantially warmer than the other
molecular clouds in W5.
}{fig:lwcas}{1.0}

The ridge is surprisingly faint in HI 21 cm emission compared to the two HII regions
(Figure \ref{fig:HIridge}) considering its 24 \um\ surface brightness.  The
integrated HI intensity from -45 to -35 \kms\ is $\sim800$ K \kms, whereas in
the HII region bubble it is around 1000 K \kms.  The CO-bright regions show
lower levels of emission similar to the ridge at 700-800 K \kms.  However, the
ridge contains no CO gas and very few young stars \citep[Figure 7 in
][]{koenig:clustered:2008}.  It is possible that the ridge contains cool HI but
has very low column-densities along the direction pointing towards the O-stars,
in which case the self-shielding is too little to prevent CO dissociation.
This ridge may therefore be an excellent location to explore the transition
from molecular to atomic gas under the influence of ionizing radiation in
conditions different from high-density photodissociation (photon-dominated)
regions.

\FigureTwo{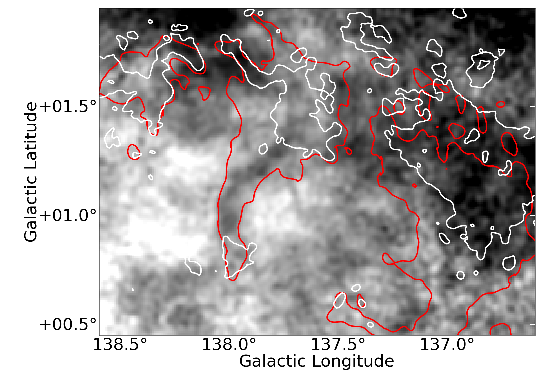}{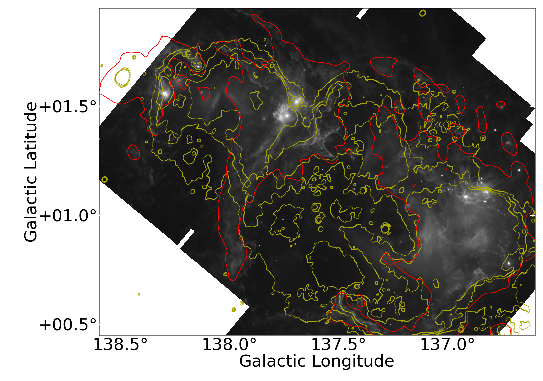}
{{\it Top:} The DRAO 21 cm HI map integrated from -45 to -35 \kms\ displayed in grayscale
from 700 (black) to 1050 (white) K \kms\ with IRAS 100 \um\ contours (red, 40 MJy sr$^{-1}$) and
\twelveco\ 1-0 contours integrated over the same range (white, 4 K \kms)
overlaid.  The ridge of IRAS 100 \um\ emission at $\ell=138.0$ coincides with a
relative lack of HI emission at these velocities, suggesting either that there
is less or colder gas along the ridge.  {\it Bottom:} The Spitzer 24 \um\ map
with 21 cm continuum contours at 6, 8, and 10 MJy sr$^{-1}$ overlaid.  The IRAS contours
are also overlaid to provide a reference for comparing the two figures and to demonstrate
that the HII region abuts the cold-HI area.  The moderate
excess of continuum emission implies a somewhat higher electron density along the 
line of sight through the ridge.}
{fig:HIridge}

We examine Outflow 20 as a possible case for pressure-driven implosion
(radiation, RDI, or gas pressure, PDI) by examining the relative timescales of the outflow
driving source and the HII-region-driven compression front.  A typical
molecular outflow source (Class 0 or I) has a lifetime of $\sim5\ee{5}$ years
\citep{Evans2009}.   Given that there is an active outflow at the head of this
cloud, we use 0.5 MYr as an upper limit.  The approximate distance from this
source to the cloud front behind it is $\sim 3.3$ pc.  If we assume the cloud
front has been pushed at a constant speed $v\leq c_{II} \approx 10 \kms$,  we
derive a lower limit on its age of 0.3 MYr.  While these limits allow for the
protostar to be older than the compression front by up to 0.2 MYr, it is likely
that the compression front moved more slowly (e.g., 3 \kms\ if it was pushed
by a D-type shock front) and that the protostar is not yet at the end of its
lifetime - it is very plausible that this soure was born in a radiation-driven
implosion.

\subsection{Southern Pillars}
\label{sec:pillars}
There are 3 cometary clouds that resemble the ``elephant trunk'' nebula in IC
1396 (Figure \ref{fig:comets}).  Each of these pillars contains evidence of at
least one outflow in the head of the cloud (see the supplementary materials, outflows
16-19 and 38)
These pillars are low-mass and isolated; there is no other outflow activity in
southern W5.  However, because of the bright illumination on their northern
edges and robust star formation tracers, these objects present a reasonable 
case for triggered star formation by the RDI mechanism.

The kinematics of these cometary clouds suggest that they have been pushed in
different directions by the HII region (Figure \ref{fig:comets}).  The central
cometary cloud (Figure \ref{fig:comets}b) has two tails.  The southwest tail
emission peaks around -39.5 \kms\ and the southeast tail peaks at -41.5 \kms,
while the head is peaked at an intermediate -40.5 \kms.  These velocity shifts
suggest that the gas is being accelerated perpendicular to the head-tail axis
and that the southeast tail is on the back side of the cometary head, while the
southwest tail is on the front side.  The expanding HII region is crushing this
head-tail system.

The southeast cometary cloud (Figure \ref{fig:comets}a) peaks at -35.0 \kms.
There are no clearly-separated CO tails as in the central cloud, but there is a
velocity shift across the tail, in which the west (right) side is blueshifted
compared to the east (left) side, which is the opposite sense from the central
cometary cloud.

The southwest cometary cloud (Figure \ref{fig:comets}c) peaks at -40.3 \kms\
and has weakly defined tails similar to the central cloud.  Both of its tails
are at approximately the same velocity (-42.5 \kms).

The kinematics of these tails provide some hints of their 3D structure and
location in the cloud.  Future study to compare the many cometary flows in W5
to physical models and simulations is warranted.  Since these flows are likely
at different locations along the line of sight (as required for their different
velocities), analysis of their ionized edges may allow for more precise
determination of the full 3D structure of the clouds relative to their ionizing
sources.

\Figure{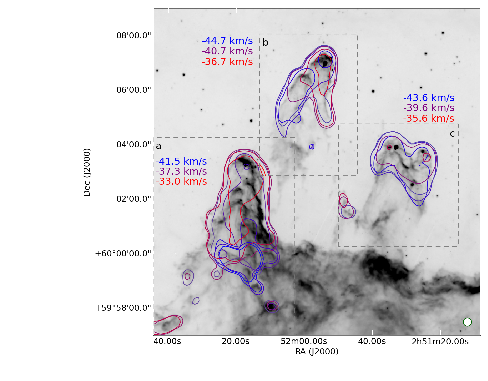}
{CO 3-2 contours overlaid on the Spitzer 8 \um\ image of the W5S cometary
clouds described in Section \ref{sec:pillars}.  Contours are color-coded by velocity
and shown for 0.84 \kms\ channels at levels of 1 K (a, b) and 0.5 K (c).  The
velocity ranges plotted are (a) -41.5 to -33.0 \kms (b) -44.7 to -36.7 \kms\
(c) -43.6 to -35.6 \kms.  The labels show the minimum, maxmimum, and middle
velocities to guide the eye.  The grey boxes indicate the regions selected for
CO contours; while there is CO emission associated with the southern 8 \um\
emission, we do not display it here. The velocity gradients are discussed in
Section \ref{sec:pillars}.
}
{fig:comets}{1.0}

\subsection{W5 Southeast}
The region identified as W5SE has very little star formation activity despite
having significant molecular gas (M$_{\thirteenco} \sim 800$\msun).  While
there are two outflows and two Class I objects \citep{koenig:clustered:2008} in the
southeast of the two clumps ($\ell=138.15,b=0.77$, Figure \ref{fig:w5SE}), the main clump
($\ell=138.0,b=0.8$) has no detected outflows.  The CO emission is particularly
clumpy in this region, with many independent, unresolved clumps both in
position and velocity.  In the 8 and 24 micron Spitzer images, it is clear that
these clouds are illuminated from the northwest.  This region represents a case
in which the expanding HII region has impacted molecular gas but has not
triggered additional star formation.  The high clump-to-clump velocity
dispersion observed in this region may be analogous to the W5S cometary clouds
(Section \ref{sec:pillars}) but without condensed clumps around which to form
cometary clouds.

\Figure{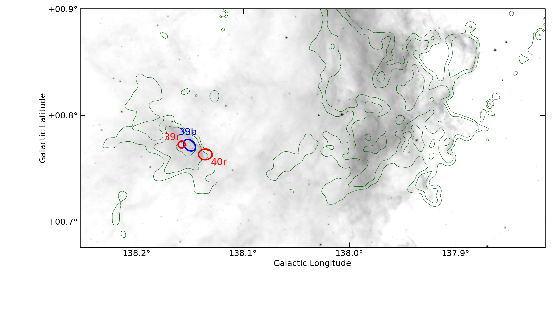}
{Small scale map of the W5 SE region showing the star-forming clump containing
outflows 39 and 40 and the non-star-forming clump at $\ell=138.0,b=0.8$. 
CO 3-2 contours integrated from -60 to -20 \kms\ are displayed at levels
3,7.2,17.3,41.6, and 100 K \kms.
}{fig:w5SE}{1.0}

\subsection{W5 Southwest}
There is an isolated clump associated with outflows in the southwest part of W5
(Figure \ref{fig:w5SW}) at $v_{LSR} \sim -45~\kms$.  While this clump is likely
to be associated with the W5 region, it shows little evidence of interaction
with the HII region.  If it is eventually impacted by the expanding ionization
front (i.e. if it is within the W5 complex), this clump will be an example of
``revealed'', not triggered, star formation.  

The other source in W5SW is a cometary cloud with a blueshifted head and
redshifted tail (Figure \ref{fig:swcomet}; Outflow 13).  The head contains a
redshifted outflow; no blueshifted counterpart was detected (the velocity
gradient displayed in Figure \ref{fig:swcomet} is smaller than the outflow
velocity and is also redshifted away from the head).  The lack of a blueshifted
counterpart may be because the flow is blowing into ionized gas where the CO is
dissociated.

Because of its evident interaction with the HII region, this source is an
interesting candidate for a non-protostellar outflow impersonator.  However,
because the head is blueshifted relative to the tail, we can infer that the
head has been accelerated towards us by pressure from the HII region, implying
that it is in the foreground of the cloud.  Given this geometry, a
radiation-driven flow would appear blueshifted, not redshifted, as the detected
flow is.  Additionally, the outflow is seen as fast as 7.5 \kms\ redshifted
from the cloud, which is a factor of 2 too fast to be driven by radiation in a
standard D-type shock.  Finally, the outflow velocity is much greater than seen
in a simulation of a cometary cloud by \citet{Gritschneder2010}, while the
head-to-tail velocity gradient is comparable.

\Figure{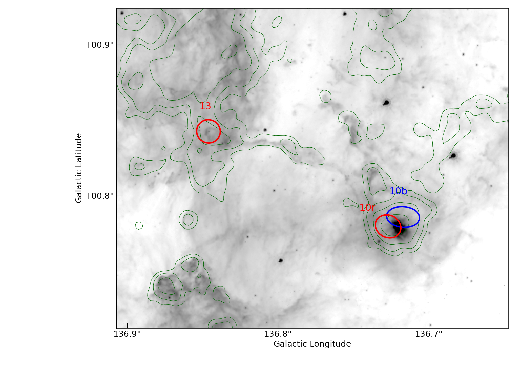}
{Small scale map of the W5 SW region plotted with CO 3-2 contours integrated
from -60 to -20 \kms\ at levels 3,7.2,17.3,41.6, and 100 K \kms. Outflow 13 is at the head of a 
cometary cloud (Figure \ref{fig:swcomet}) and therefore has clearly been
affected by the expanding HII region, but the region including bipolar Outflow
10 shows no evidence of interaction with the HII region. 
}{fig:w5SW}{1.0}

\Figure{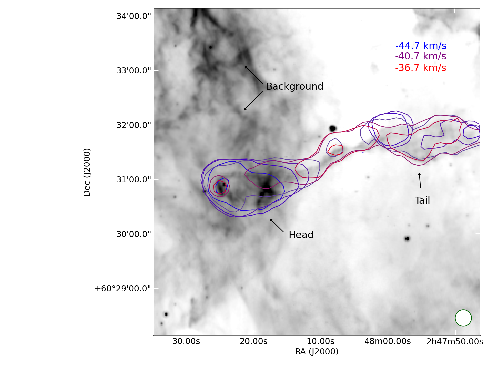}
{The cometary cloud in the W5 Southwest region (Outflow 13).  Contours are
shown at 1 K for 0.84 \kms\ wide channels from -37.2 \kms\ (blue) to -30.5
\kms\ (red).  The head is clearly blueshifted relative to the tail and contains
a spatially unresolved redshifted outflow.}
{fig:swcomet}{1.0}

\subsection{W5 West / IC 1848}
\label{sec:i02459}
There is a bright infrared source seen in the center of W5W (IRAS 02459+6029;
Figure \ref{fig:w5w}), but the nearest CO outflow lobe is $\approx1$ pc away.
The nondetection may be due to confusion in this area: there are two layers of
CO gas separated by $\sim$5 \kms, so low-velocity outflow detection is more
difficult. 
Unlike the rest of the W5 complex, this region appears to have multiple
independent confusing components along the line of sight (Figure
\ref{fig:regionspectra}), and therefore the CO data provide much less
useful physical information (multiple components are also observed in the
\thirteenco\ data, ruling out self-absorption as the cause of the multiple
components).

\Figure{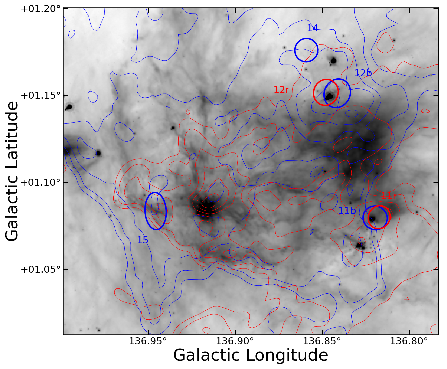}
{Small scale map of the W5 W region.  The IRAC 8 \um\ image is displayed in
inverted log scale from \lowirac\ to \highirac\ MJy \persr.  Contours of the CO
3-2 cube integrated from -50 to -38 \kms\ (blue) and -38 to -26 \kms\ (red) are
overlaid at levels 5,10,20,30,40,50,60 K \kms\ $\sigma\approx0.5$ K \kms.  The
lack of outflow detections is partly explained by the two spatially overlapping
clouds that are adjacent in velocity.
}{fig:w5w}{1.0}

\subsection{W5 NW}
The northwest cluster  containing outflows 1-8 is at a
slightly different velocity ($\sim-35~\kms$) than the majority of the W5 cloud
complex ($\sim-38~\kms$; Figure \ref{fig:w5pv}), but it shares contiguous
emission with the neighboring W5W region.  
It contains many outflows and therefore is actively forming stars  (Figure
\ref{fig:w5nw}).  However, this cluster exhibits much lower CO brightness
temperatures and weaker Spitzer 8 \um\ emission than the ``bright-rimmed
clouds'' seen near the W5 O-stars. We therefore conclude that the region has
not been directly impacted by any photoionizing radiation from the W5 O-stars.

The lack of interaction with the W5 O-stars implies that the star formation in
this region, though quite vigorous, has not been directly triggered.  Therefore
not all of the current generation of star formation in W5 has been triggered on
small or intermediate scales (e.g., radiation-driven implosion).  Even the
``collect and collapse'' scenario seems unlikely here, as the region with the
most outflows also displays some of the smoothest morphology (Figures
\ref{fig:outflows_on_co32} and \ref{fig:w5nw}); in ``collect and collapse'' the
expansion of an HII region leads to clumping and fragmentation, and the spaces
between the clumps should be relatively cleared out.

\Figure{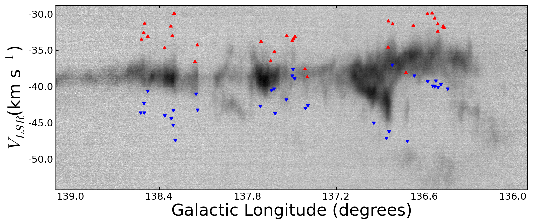}
{Integrated longitude-velocity diagram of the W5 complex from $b=0.25$ to
$b=2.15$ in \twelveco\ 1-0 from the FCRAO OGS.  The W5NW region is seen at a
distinct average velocity around $\ell=136.5$, $v_{LSR}=-34$ \kms.  The red and
blue triangles mark the longitude-velocity locations of the detected outflows.
In all cases, they mark the low-velocity start of the outflow.}
{fig:w5pv}{1.0}

\Figure{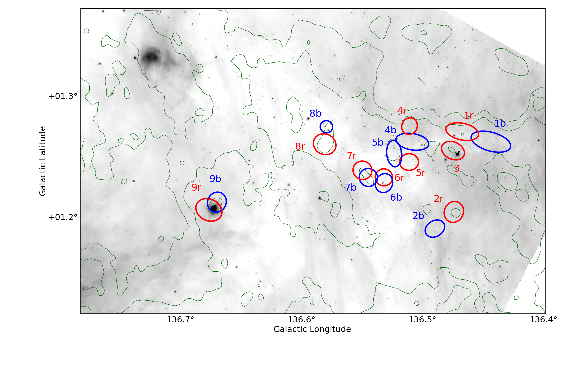}
{Small scale map of the W5 NW region plotted with CO 3-2 contours integrated
from -60 to -20 \kms\ at levels 3,7.2,17.3,41.6, and 100 K \kms. Despite its
distance from the W5 O-stars, $D_{proj}\approx20$ pc, this cluster is the most
active site of star formation in the complex as measured by outflow activity.}
{fig:w5nw}{1.0}

\section{Discussion}

\subsection{Comparison to other outflows}
\label{sec:comparison}
The outflow properties we derive are similar to those in the B0-star forming
clump IRAS 05358+3543 \citep[$M\approx600\msun$][]{Ginsburg2009}, in which CO
3-2 and 2-1 were used to derive outflow masses in the range 0.01-0.09 \msun.
However, some significantly larger outflows, up to 1.6 pc in one direction were
detected, while the largest resolved outflow in our survey was only 0.8 pc (one
direction).  

As noted in Section \ref{sec:percompare}, 
the total molecular mass in W5 is larger than
Perseus, $M_{W5}\sim 4.5\ee{4} \msun$ while $M_{Perseus}\sim10^4\msun$
\citep{bally-perseus2008}.  The length distribution of outflows (Figure
\ref{fig:lengthhist}) is strikingly similar, while other physical properties
have substantially different mean values with or without correction factors
included.

The W5NW region is more directly comparable to Perseus, with a total mass of
$\sim1.5\ee{4}$ \msun\ (Table \ref{tab:regionspectra}) and a similar size.  In
Figure \ref{fig:regionboxes_on_CO}, we show both the W5NW region, which
contains all of the identified outflows, and the W5NWpc region, which is a
larger area intended to be directly comparable in both mass and spatial scale
to the Perseus molecular cloud.  The W5NWpc region contains more than an order of
magnitude more turbulent energy than the Perseus complex \citep[$E_{turb,Per} =
1.6\ee{46}$ ergs,][]{arce2010} despite its similar mass.  Even the smaller W5NW
region has $\sim5\times$ more turbulent energy than the Perseus complex,
largely because of the greater average line width ($\sigma_{FWHM,W5NW}\approx
3.5$ \kms).  As with the whole of W5, there is far too much turbulent energy in
W5NW to be provided by outflows alone, implying the presence of another driver
of turbulence.

Figure \ref{fig:w5nwpercompare} shows the W5NWpc region and Perseus molecular cloud
on the same scale, though in two different emission lines.  The Perseus cloud
contains many more outflows and candidates (70 in Perseus vs. 13 in W5NWpc)
despite a much larger physical area surveyed in W5.  While it is likely that
many of the W5W outflows will break apart into multiple flows at higher
resolution, it does not seem likely that each would break apart into 5 flows,
as would be required to bring the numbers into agreement.  Since the highest
density of outflows in Perseus is in the NGC 1333 cloud, it may be that there
is no equivalently evolved region in W5NWpc.  The W5W region may be comparably
massive, but it is also confused and strongly interacting with the W5 HII
region - either star formation is suppressed in this region, or outflows are
rendered undetectable.  In the latter case, the most likely mechanisms for
hiding outflows are molecular dissociation by ionizing radiation and velocity
confusion.

Another possibility highlighted in Figure \ref{fig:w5nwpercompare} is that the
W5NW region is interacting with the W4 bubble.  The cloud in the top right of
Figure \ref{fig:w5nwpercompare} is somewhat cometary, has higher peak
brightness temperature, and is at a slightly different velocity (-45 \kms) than
W5NW.  The velocity difference of $\sim8$ \kms could simply be two clouds
physically unassociated along the line of sight, or could indicate the presence
of another expanding bubble pushing two sheets of gas away from each other.
Either way, the northwest portion of the W5NW region is clumpier than the
area in which the outflows were detected, and it includes no outflow detections.

\FigureTwo{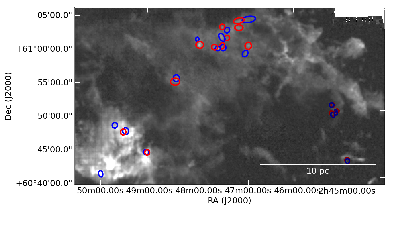}{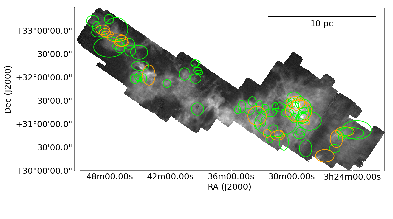}
{(a) An integrated CO 3-2 image of the W5W/NW region with ellipses overlaid
displaying the locations and sizes of outflows.  The dark red and blue ellipses
in the lower right are associated with outer-arm outflows.  W5W is the bottom-left,
CO-bright region.  W5NW is the top-center region containing the cluster of outflows.
(b) An integrated CO 1-0 image of the Perseus molecular cloud from the COMPLETE
survey \citep{arce2010}.  Note that the spatial scale is identical to that of
(a) assuming that W5 is 8 times more distant than Perseus.  The green ellipses
represent \citet{arce2010} CPOCs while the orange represent known outflows from
the same paper.}
{fig:w5nwpercompare}

\subsection{Star Formation Activity}
\label{sec:sfactivity}

CO outflows are an excellent tracer of ongoing embedded star formation
\citep[e.g.][]{Shu1987}.  We use the locations of newly discovered outflows to
qualitatively describe the star formation activity within the W5 complex and
evaluate the hypothesis that star formation has been triggered on small or
intermediate scales.

Class 0/I objects are nearly always associated with outflows in nearby
star-forming regions \citep[e.g. Perseus][]{curtis2010,hatchell2007}.  However,
\citet{koenig:clustered:2008} detected 171 Class I sources in W5 using Spitzer
photometry.  Since our detection threshold for outflow appears to be similar to
that in Perseus (Section \ref{sec:percompare}), the lower number of outflow
detections is surprising, especially considering that some of the detected
outflows are outside the Spitzer-MIPS field (MIPS detections are required for
Class I objects, and flows 1-4 are outside that range) or are in the outer arm
(flows 39-54).  Additionally, we should detect outflows from Class 0
objects that would not be identified by Spitzer colors.

There are a number of explanations for our detection deficiency.  The Class I
objects detected within the HII region ``bubble'' most likely have outflows in
which the CO is dissociated similar to jet systems in Orion \citep[e.g.
HH46/47, a pc-scale flow in which CO is only visible very near the
protostar;][]{Chernin1991,Stanke1999}.  This hypothesis can be tested by
searching for optical and infrared jets associated with these objects, which
presumably have lower mass envelopes and therefore less extinction than
typical Class I objects.  Additionally, there are many outflow systems that are
are likely to be associated with clusters of outflows rather than individual
outflows as demonstrated in Section \ref{sec:percompare}, where we were able to identify
fewer outflows when `observing' the Perseus objects at a greater distance.  There
are 24 sources in the \citet{koenig:clustered:2008} Class I catalog within
15\arcsec\ (one JCMT beam at 345 GHz) of another, and in many cases there are
multiple \citet{koenig:clustered:2008} Class I sources within the contours of a
single outflow system.

\subsection{Evaluating Triggering}
In the previous section, we discussed in detail the relationship between each
sub-region and the HII region.  Some regions are observed to be star-forming
but not interacting with the HII region (W5NW, Sh 2-201), while others are
interacting with the HII region but show no evidence or reduced evidence of
star formation (W5SE, W5W, LW Cas).  At the very least, there is significant
complexity in the triggering mechanisms, and no one mechanism or size scale is
dominant.  If we were to trust outflows as unbiased tracers of star formation,
we might conclude that the majority of star formation in W5 is untriggered
(spontaneous), but such a conclusion is unreliable because both radiatively
triggered star formation and ``revealed'' star formation may not exhibit
molecular outflows (although ionized atomic outflows should still be visible
around young stars formed through these scenarios).

In Section \ref{sec:w5ridge}, we analyzed a particular case in which the RDI
mechanism could plausibly have crushed a cloud to create the observed
protostar.  
It is not possible to determine whether interaction with the HII region was a 
necessary precondition for the star's formation, but it at least accelerated
the process.  The other cometary clouds share this property,
but in total there are only 5 cometary clouds with detected outflows at their
tips, indicating that this mechanism is not the dominant driver of star
formation in W5.

The `collect and collapse' scenario might naively be expected to produce an
excess of young stars at the interaction front between the HII region and the
molecular cloud.  However, because such interactions naturally tend to form
instabilities, this scenario produces cloud morphologies indistinguishable from
those of RDI.  There is not an obvious excess of sources associated with cloud
edges over those deep within the clouds (e.g., Figure
\ref{fig:outflows_on_co32}).  We therefore cannot provide any direct evidence
for this triggering scenario.

The overall picture of W5 is of two concurrent episodes of massive-star
formation that have lead to adjacent blown-out bubbles.  Despite the added
external pressure along the central ridge, it is relatively deficient in both
star formation activity and dense gas, perhaps because of heating by the strong
ionizing radiation field.  The lack of star formation along that central ridge
implies that much of the gas was squeezed and heated, but it was not crushed
into gravitationally unstable fragments.  While some star formation may have
been triggered in W5, there is strong evidence for pre-existing star formation
being at least a comparable, if not the dominant, mechanism of star formation
in the complex.

\section{Outflow systems beyond W5}
Fifteen outflows were detected at velocities inconsistent with the local W5
cloud velocities.  Of these, 8 are consistent with Perseus arm velocities
($v_{LSR} > -55$ \kms) and could be associated with different clouds within the
same spiral arm.  The other 7 have central velocities $v_{LSR} < -55$ \kms\ and
are associated with the outer arm identified in previous surveys
\citep[e.g.][]{Digel1996}.  The properties of these outflows are given in Tables
\ref{tab:outeroutflows} and \ref{tab:outeroutflowsderived}; the distances listed are kinematic distances assuming
$R_0=8.4$ kpc and $v_0=254$~\kms\ \citep{Reid2009}.

Of these outflows, all but one are within 2\arcmin\ of an IRAS point source.
Outflow 54 is the most distant in our survey at a kinematic distance $d=7.5$
kpc ($v_{lsr}=-75.6$ \kms) and galactocentric distance $D_G = 14.7$ kpc.  It
has no known associations in the literature.

Outflows 41 - 44 are associated with a cloud at $v_{LSR}\sim -62$ \kms\ known
in the literature as LDN 1375 and associated with IRAS 02413+6037.  Outflows 53
and 55 are at a similar velocity and associated with IRAS 02598+6008 and IRAS
02425+6000 respectively.  All of these sources lie roughly on the periphery of
the W5 complex.

Outflows 45 - 52 are associated with a string of IRAS sources and HII regions
to the north of W5 and have velocities in the range $-55 < v_{LSR} < -45$.
They therefore could be in the Perseus arm but are clearly unassociated with
the W5 complex.  Outflows 45 and 46 are associated with IRAS 02435+6144 and
they may also be associated with the Sharpless HII region Sh 2-194.  Outflows
47 and 48 are associated with IRAS 02461+6147, also known as AFGL 5085.
Outflows 49 and 50 are nearby but not necessarily associated with IRAS
02475+6156, and may be associated with Sh 2-196.  Outflows 51 and 52 are
associated with IRAS 02541+6208.

\section{Conclusions}

We have identified \nwfive\ molecular outflow candidates in the W5 star forming
region and an additional \nouter\ outflows spatially coincident but located in
the outer arm of the Galaxy.  

\begin{itemize}
    \item The majority of the CO clouds in the W5 complex are forming stars.
      Star formation is not limited to cloud edges around the HII region.
      Because star formation activity is observed outside of the region of
      influence of the W5 O-stars, it is apparent that direct triggering by
      massive star feedback is not responsible for all of the star formation in
      W5.
    \item The W5 complex is seen nearly face-on as evidenced by a strict upper
      limit on the CO column through the center of the HII-region bubbles.  It
      is therefore an excellent region to study massive star feedback and
      revealed and triggered star formation.
    \item Outflows contribute negligibly to the turbulent energy of molecular
      clouds in the W5 complex.  This result is unsurprising near an HII
      region, but supports the idea that massive star forming regions are
      qualitatively different from low-mass star-forming regions in which the
      observed turbulence could be driven by outflow feedback.  Even in regions
      far separated from the O-stars, there is more turbulence and less energy
      injection from outflows than in, e.g., Perseus.
    \item Despite detecting a significant number of powerful outflows, the
      total outflowing mass detected in this survey ($\sim 1.5$ \msun\ without
      optical depth correction, perhaps $10-20$ \msun\ when optical depth is
      considered) was somewhat smaller than in Perseus, a low to intermediate
      mass star forming region with $\sim 1/6$ the molecular mass of W5. 
    \item The low mass measured may be partly because the CO 3-2 line is
      sub-thermally excited in outflows.  Therefore, while CO 3-2 is an
      excellent tracer of outflows for detection, it does not serve as a
      `calorimeter' in the same capacity as CO 1-0.
    \item Even considering excitation and optical depth corrections, it is
      likely that the mass of outflows in W5 is less than would be expected
      from a simple extrapolation from Perseus based on cloud mass. CO is
      likely to be photodissociated in the outflows when they reach the HII
      region, accounting for the deficiency around the HII region edges.
      However, in areas unaffected by the W5 O-stars such as W5NW, the
      deficiency may be because the greater turbulence in the W5 clouds
      suppresses star formation or hides outflows.
    \item Velocity gradients across the tails of many cometary clouds have been
      observed, hinting at their geometry and confirming that the outflows seen
      from their heads must be generated by protostars within.
    \item Outflows have been detected in the Outer Arm at galactocentric
      distances $\gtrsim12$ kpc.  These represent some of the highest
      galactocentric distance star forming regions discovered to date.
\end{itemize}

\section{Acknowledgements}
We thank the two anonymous referees for their assistance in refining this
document.  We thank Devin Silvia for a careful proofread of the text. This work
has made use of the APLpy plotting package
(\url{http://aplpy.sourceforge.net}), the pyregion package
(\url{http://leejjoon.github.com/pyregion/}), the agpy code package
(\url{http://code.google.com/p/agpy/}) , IPAC's Montage
(\url{http://montage.ipac.caltech.edu/}), the DS9 visualization tool
(\url{http://hea-www.harvard.edu/RD/ds9/}), the pyspeckit spectrosopic analysis
toolkit (\url{http://pyspeckit.bitbucket.org}), and the STARLINK package
(\url{http://starlink.jach.hawaii.edu/}).  IRAS data was acquired through IRSA
at IPAC (\url{http://irsa.ipac.caltech.edu/}).  DRAO 21 cm data was acquired
from the Canadian Astronomical Data Center
(\url{http://cadcwww.hia.nrc.ca/cgps/}).  The authors are supported by the
National Science Foundation through NSF grant AST-0708403.  This research has
made use of the SIMBAD database, operated at CDS, Strasbourg, France

{\it Facilities:} JCMT, VLA

\clearpage
\onecolumn
\LongTable{ccccccccccc}{CO 3-2 Outflow Measured Properties}
{{Outflow} & {Latitude} & {Longitude} & {Ellipse} & {Ellipse} & {Ellipse} & {Velocity} & {Velocity} &  & {$\int T_A^* dv$} & {Bipolar?  \tablenotemark{a}}\\
{Number} &  &  & {Major} & {Minor} & {PA} & {center} & {min} & {max} &  & \\
 &  &  & {\arcsec} & {\arcsec} & {\degree} & {(\kms)} & {(\kms)} & {(\kms)} & {(K \kms)} & \\}
{tab:outflows}
{
1b & 136.4437 & 1.2622 & 60 & 27 & 342 & -36.1 & -47.6 & -40.3 & 1.0 & yc\\
1r & 136.4674 & 1.2705 & 49 & 24 & 346 & -36.1 & -31.9 & -23.4 & 1.5 & yc\\
2b & 136.4899 & 1.1904 & 30 & 23 & 299 & -35.7 & -48.0 & -39.7 & 0.7 & yc\\
2r & 136.4743 & 1.2042 & 31 & 28 & 332 & -35.7 & -31.7 & -23.0 & 1.3 & yc\\
3 & 136.475 & 1.2548 & 35 & 25 & 332 & -31.8 & -31.8 & -26.8 & 1.3 & n\\
4b & 136.5038 & 1.2623 & 26 & 22 & 35 & -36.2 & -44.1 & -40.1 & 0.8 & yu\\
4r & 136.5109 & 1.2751 & 25 & 22 & 332 & -36.2 & -32.4 & -28.6 & 0.9 & yu\\
5r & 136.5126 & 1.2453 & 24 & 22 & 10 & -35.3 & -31.4 & -28.8 & 0.8 & yu\\
5b & 136.5236 & 1.2524 & 39 & 22 & 3 & -35.3 & -45.0 & -39.2 & 1.4 & yu\\
6b & 136.532 & 1.228 & 28 & 25 & 332 & -35.3 & -44.8 & -40.0 & 0.4 & yc\\
6r & 136.5327 & 1.2333 & 28 & 20 & 318 & -35.3 & -30.6 & -24.0 & 1.0 & yc\\
7b & 136.5453 & 1.2318 & 24 & 19 & 332 & -34.9 & -47.5 & -39.9 & 1.7 & yc\\
7r & 136.5506 & 1.2383 & 27 & 23 & 314 & -34.9 & -29.9 & -22.7 & 1.3 & yc\\
8b & 136.5799 & 1.2755 & 18 & 14 & 332 & -34.5 & -41.5 & -39.3 & 0.6 & yc\\
8r & 136.581 & 1.2601 & 34 & 30 & 332 & -34.5 & -29.6 & -23.9 & 1.4 & yc\\
9b & 136.67 & 1.2123 & 30 & 27 & 332 & -35.0 & -44.5 & -38.5 & 1.4 & yc\\
9r & 136.6766 & 1.2059 & 40 & 31 & 332 & -35.0 & -31.6 & -26.7 & 0.3 & yc\\
10b & 136.7172 & 0.7859 & 39 & 24 & 353 & -42.8 & -52.6 & -47.5 & 3.3 & yc\\
10r & 136.7271 & 0.7797 & 31 & 26 & 332 & -42.8 & -38.1 & -33.1 & 4.1 & yc\\
11b & 136.8195 & 1.082 & 25 & 24 & 331 & -34.2 & -40.7 & -37.0 & 3.1 & yc\\
11r & 136.8173 & 1.0799 & 24 & 22 & 331 & -34.2 & -31.4 & -20.4 & 1.5 & yc\\
12b & 136.8414 & 1.1512 & 30 & 26 & 332 & -40.4 & -53.3 & -46.2 & 1.5 & yc\\
12r & 136.8479 & 1.1517 & 27 & 25 & 332 & -40.4 & -34.6 & -30.1 & 0.9 & yc\\
13 & 136.8461 & 0.8426 & 28 & 27 & 332 & -31.0 & -31.0 & -23.5 & 1.0 & n\\
14 & 136.8591 & 1.176 & 24 & 23 & 332 & -47.1 & -54.5 & -47.1 & 0.8 & n\\
15 & 136.9443 & 1.0841 & 28 & 18 & 348 & -45.0 & -55.0 & -45.0 & 3.1 & n\\
16b & 137.3929 & 0.5977 & 23 & 18 & 333 & -40.7 & -47.0 & -42.6 & 0.7 & yu\\
16r & 137.3981 & 0.6121 & 22 & 19 & 357 & -40.7 & -38.7 & -35.2 & 1.9 & yu\\
17b & 137.4084 & 0.6762 & 20 & 18 & 293 & -40.3 & -57.9 & -43.0 & 2.3 & yc\\
17r & 137.412 & 0.6775 & 20 & 18 & 308 & -40.3 & -37.6 & -30.4 & 1.1 & yc\\
18b & 137.4925 & 0.6289 & 16 & 15 & 333 & -35.5 & -39.2 & -37.6 & 1.1 & yc\\
18r & 137.4908 & 0.6292 & 18 & 17 & 307 & -35.5 & -33.4 & -31.0 & 2.0 & yc\\
19b & 137.4815 & 0.6409 & 20 & 17 & 1 & -36.0 & -41.9 & -38.9 & 1.3 & yc\\
19r & 137.4798 & 0.6404 & 20 & 16 & 301 & -36.0 & -33.1 & -25.9 & 0.7 & yc\\
20r & 137.5368 & 1.2792 & 24 & 21 & 332 & -37.4 & -33.0 & -22.5 & 5.2 & yc\\
20b & 137.539 & 1.279 & 27 & 23 & 17 & -37.4 & -52.0 & -41.8 & 3.4 & yc\\
21b & 137.6152 & 1.3543 & 31 & 28 & 322 & -39.5 & -52.0 & -43.7 & 4.5 & yc\\
21r & 137.6169 & 1.3585 & 31 & 18 & 4 & -39.5 & -35.2 & -30.0 & 1.2 & yc\\
22 & 137.6213 & 1.506 & 27 & 21 & 293 & -40.3 & -46.0 & -40.3 & 2.1 & n\\
23b & 137.6389 & 1.5251 & 21 & 14 & 331 & -38.5 & -42.5 & -40.5 & 1.6 & yc\\
23r & 137.6449 & 1.5194 & 19 & 12 & 331 & -38.5 & -36.5 & -32.0 & 1.9 & yc\\
24r & 137.7094 & 1.4824 & 20 & 20 & 331 & -38.2 & -33.8 & -25.4 & 4.2 & yc\\
24b & 137.7146 & 1.4809 & 25 & 19 & 292 & -38.2 & -50.0 & -42.7 & 4.4 & yc\\
25b & 138.1398 & 1.6858 & 39 & 26 & 282 & -38.8 & -49.5 & -43.2 & 0.6 & yc\\
25r & 138.142 & 1.6884 & 43 & 35 & 11 & -38.8 & -34.3 & -27.5 & 1.7 & yc\\
26b & 138.2913 & 1.5538 & 29 & 29 & 355 & -38.7 & -52.0 & -47.4 & 1.2 & yc\\
26r & 138.2966 & 1.5564 & 28 & 28 & 330 & -38.7 & -30.0 & -20.0 & 4.2 & yc\\
27 & 138.3017 & 1.5689 & 26 & 25 & 330 & -30.0 & -30.0 & -22.0 & 1.8 & n\\
28 & 138.3042 & 1.5437 & 20 & 19 & 330 & -43.3 & -46.1 & -43.3 & 1.4 & n\\
29 & 138.3053 & 1.5537 & 22 & 20 & 330 & -45.3 & -51.6 & -45.3 & 2.5 & n\\
30 & 138.3115 & 1.5443 & 26 & 26 & 330 & -33.0 & -33.0 & -29.2 & 1.2 & n\\
31 & 138.3184 & 1.5566 & 26 & 25 & 330 & -44.4 & -49.1 & -44.4 & 1.1 & n\\
32 & 138.3213 & 1.5658 & 27 & 27 & 330 & -31.7 & -31.7 & -27.0 & 1.4 & n\\
33b & 138.3618 & 1.5073 & 28 & 26 & 330 & -39.4 & -49.5 & -44.0 & 1.3 & yc\\
33r & 138.3642 & 1.4959 & 29 & 21 & 330 & -39.4 & -34.7 & -25.8 & 2.0 & yc\\
34r & 138.4779 & 1.6137 & 22 & 21 & 330 & -36.9 & -33.1 & -29.1 & 0.5 & yc\\
34b & 138.4768 & 1.6142 & 21 & 20 & 330 & -36.9 & -43.6 & -40.6 & 0.8 & yc\\
35r & 138.4998 & 1.6496 & 22 & 20 & 4 & -37.5 & -31.3 & -24.1 & 1.4 & yc\\
35b & 138.5021 & 1.6458 & 23 & 21 & 330 & -37.5 & -49.5 & -43.6 & 1.3 & yc\\
36b & 138.5034 & 1.6654 & 35 & 26 & 5 & -37.5 & -50.4 & -42.3 & 1.2 & yc\\
36r & 138.5061 & 1.6576 & 22 & 21 & 330 & -37.5 & -32.6 & -26.7 & 1.4 & yc\\
37r & 138.5208 & 1.6618 & 27 & 22 & 330 & -38.5 & -33.5 & -31.4 & 0.6 & yc\\
37b & 138.5241 & 1.6667 & 23 & 23 & 18 & -38.5 & -47.0 & -43.6 & 0.6 & yc\\
38b & 137.4983 & 0.6062 & 16 & 15 & 333 & -36.1 & -39.2 & -38.5 & 0.8 & yc\\
38r & 137.4977 & 0.6055 & 15 & 14 & 307 & -36.1 & -33.7 & -32.5 & 0.5 & yc\\
39b & 138.1506 & 0.7724 & 23 & 16 & 321 & -38.8 & -45.3 & -41.0 & 2.0 & yc\\
39r & 138.1591 & 0.7713 & 17 & 13 & 304 & -38.8 & -36.6 & -34.7 & 0.7 & yc\\
40 & 138.1356 & 0.7634 & 22 & 18 & 4 & -36.0 & -36.0 & -27.6 & 2.2 & n\\
}{\multicolumn{11}{l}{Measured properties of the outflows.} \\ 
\multicolumn{11}{l}{$^a$ Is the outflow part of a bipolar pair?  yc = yes, confident; yu = yes, uncertain; n = no} \\ }
{}
{11}

\clearpage

\LongTable{cccccc}{CO 3-2 Outflow Derived Properties}
{{Outflow} & {Mass} & {Momentum} & {Energy} & {Dynamical} & {Momentum}\\
{Number} &  &  &  & {Age} & {Flux}\\
 & {(\msun)} & {(\msun \kms)} & {(10$^{42}$ ergs)} & {(10$^4$ years)} & {$10^{-6}$ \msun}\\
 &  &  &  &  & {\kms yr$^{-1}$}\\}
{tab:outflowsderived}
{
1b & 0.034 & 0.26 & 21.1 & 7.0 & 7.2\\
1r & 0.04 & 0.24 & 17.3 & 7.0 & 7.2\\
2b & 0.011 & 0.07 & 4.9 & 5.4 & 4.4\\
2r & 0.025 & 0.17 & 13.0 & 5.4 & 4.4\\
3 & 0.025 & 0.12 & 5.8 & - & -\\
4b & 0.01 & 0.06 & 4.0 & 7.2 & 1.5\\
4r & 0.011 & 0.04 & 1.8 & 7.2 & 1.5\\
5r & 0.01 & 0.04 & 2.0 & 4.5 & 4.0\\
5b & 0.025 & 0.14 & 8.0 & 4.5 & 4.0\\
6b & 0.007 & 0.04 & 3.0 & 1.7 & 8.1\\
6r & 0.013 & 0.09 & 6.8 & 1.7 & 8.1\\
7b & 0.017 & 0.13 & 10.5 & 2.4 & 10.9\\
7r & 0.018 & 0.13 & 9.7 & 2.4 & 10.9\\
8b & 0.003 & 0.02 & 0.9 & 4.9 & 4.5\\
8r & 0.032 & 0.2 & 13.2 & 4.9 & 4.5\\
9b & 0.025 & 0.13 & 7.2 & 3.9 & 4.2\\
9r & 0.009 & 0.04 & 1.8 & 3.9 & 4.2\\
10b & 0.068 & 0.41 & 25.7 & 3.9 & 22.0\\
10r & 0.074 & 0.45 & 28.0 & 3.9 & 22.0\\
11b & 0.042 & 0.17 & 7.2 & 0.7 & 35.3\\
11r & 0.017 & 0.09 & 5.9 & 0.7 & 35.3\\
12b & 0.026 & 0.14 & 8.7 & 1.8 & 15.2\\
12r & 0.014 & 0.13 & 12.7 & 1.8 & 15.2\\
13 & 0.016 & 0.1 & 5.8 & - & -\\
14 & 0.01 & 0.06 & 4.1 & - & -\\
15 & 0.036 & 0.24 & 17.3 & - & -\\
16b & 0.006 & 0.03 & 1.2 & 11.1 & 0.7\\
16r & 0.018 & 0.05 & 1.3 & 11.1 & 0.7\\
17b & 0.019 & 0.12 & 9.4 & 1.4 & 10.6\\
17r & 0.009 & 0.03 & 0.7 & 1.4 & 10.6\\
18b & 0.006 & 0.02 & 0.5 & 1.4 & 4.0\\
18r & 0.013 & 0.04 & 1.0 & 1.4 & 4.0\\
19b & 0.011 & 0.05 & 2.4 & 0.7 & 9.6\\
19r & 0.005 & 0.01 & 0.2 & 0.7 & 9.6\\
20r & 0.059 & 0.5 & 46.3 & 0.5 & 156.0\\
20b & 0.047 & 0.33 & 26.6 & 0.5 & 156.0\\
21b & 0.086 & 0.58 & 41.4 & 1.7 & 39.1\\
21r & 0.014 & 0.08 & 4.3 & 1.7 & 39.1\\
22 & 0.027 & 0.1 & 4.3 & - & -\\
23b & 0.011 & 0.03 & 0.9 & 4.5 & 1.3\\
23r & 0.01 & 0.03 & 1.0 & 4.5 & 1.3\\
24r & 0.037 & 0.3 & 26.1 & 1.7 & 34.1\\
24b & 0.047 & 0.28 & 18.3 & 1.7 & 34.1\\
25b & 0.014 & 0.09 & 6.8 & 1.0 & 42.8\\
25r & 0.056 & 0.35 & 23.0 & 1.0 & 42.8\\
26b & 0.023 & 0.24 & 26.1 & 1.1 & 98.3\\
26r & 0.072 & 0.85 & 106.0 & 1.1 & 98.3\\
27 & 0.026 & 0.09 & 4.5 & - & -\\
28 & 0.012 & 0.07 & 4.6 & - & -\\
29 & 0.024 & 0.06 & 2.1 & - & -\\
30 & 0.018 & 0.12 & 8.0 & - & -\\
31 & 0.016 & 0.03 & 0.7 & - & -\\
32 & 0.023 & 0.18 & 14.5 & - & -\\
33b & 0.022 & 0.14 & 10.1 & 3.0 & 11.4\\
33r & 0.026 & 0.2 & 16.1 & 3.0 & 11.4\\
34r & 0.005 & 0.03 & 1.7 & 0.7 & 8.4\\
34b & 0.007 & 0.03 & 1.2 & 0.7 & 8.4\\
35r & 0.013 & 0.12 & 11.6 & 1.3 & 18.7\\
35b & 0.014 & 0.12 & 11.0 & 1.3 & 18.7\\
36b & 0.025 & 0.19 & 15.8 & 2.7 & 10.7\\
36r & 0.014 & 0.1 & 6.8 & 2.7 & 10.7\\
37r & 0.008 & 0.04 & 1.6 & 2.1 & 4.3\\
37b & 0.007 & 0.06 & 4.2 & 2.1 & 4.3\\
38b & 0.005 & 0.01 & 0.4 & 0.9 & 2.3\\
38r & 0.002 & 0.01 & 0.1 & 0.9 & 2.3\\
39b & 0.017 & 0.07 & 2.8 & 7.5 & 1.0\\
39r & 0.004 & 0.01 & 0.3 & 7.5 & 1.0\\
40 & 0.019 & 0.08 & 3.5 & - & -\\
}{\multicolumn{6}{l}{Derived properties of the outflows in
the optically thin limit.} \\  
\multicolumn{6}{l}{Typical optical depth corrections for \twelveco 3-2 are
in the range 7-14 \citep{curtis2010}.} \\  
\multicolumn{6}{l}{The correction for velocity confusion is $\gtrsim2$ but
poorly constrained \citep{arce2010}.} \\  
\multicolumn{6}{l}{Finally, an excitation correction in the range 1-20 is
likely required as described in the Appendix.} \\
\multicolumn{6}{l}{The mass and momentum values can be multiplied by these
factors to acquire the corrected values.} \\  
\multicolumn{6}{l}{The energy is weighted more heavily towards high-velocity,
low-optical-depth gas, so the correction factor is likely to be lower.}}
{} {6}

\clearpage
\Table{cccccccccccc}{Outer Arm CO 3-2 Outflows - Measured Properties}
{{Outflow} & {Latitude} & {Longitude} & {Ellipse} & {Ellipse} & {Ellipse} & {Kinematic} & {$R_G$\tablenotemark{a}} & {Velocity} & {Velocity} &  & {$\int T_A^* dv$}\\
{Number} &  &  & {Major} & {Minor} & {PA} & {Distance} &  & {center} & {min} & {max} & \\
 &  &  & {\arcsec} & {\arcsec} & {\degree} & {(pc)} & {(pc)} & {(\kms)} &  &  & {(K \kms)}\\}
{tab:outeroutflows}
{
41r & 136.364 & 0.9606 & 25 & 18 & 2 & 5510 & 13000 & -61.8 & -59.2 & -56.5 & 0.5\\
41b & 136.3634 & 0.9568 & 23 & 17 & 353 & 5510 & 13000 & -61.8 & -71.6 & -64.3 & 3.0\\
42r & 136.3522 & 0.9786 & 20 & 14 & 2 & 5500 & 12900 & -62.1 & -59.8 & -57.6 & 0.6\\
42b & 136.3548 & 0.9798 & 20 & 19 & 332 & 5500 & 12900 & -62.1 & -67.8 & -64.4 & 0.5\\
43r & 136.3495 & 0.9612 & 17 & 15 & 63 & 5510 & 13000 & -61.8 & -59.0 & -56.1 & 0.8\\
43b & 136.353 & 0.9621 & 12 & 12 & 333 & 5510 & 13000 & -61.8 & -66.3 & -64.6 & 1.0\\
44r & 136.3554 & 0.9576 & 13 & 13 & 23 & 5500 & 12900 & -61.8 & -59.0 & -55.4 & 2.1\\
44b & 136.3545 & 0.9567 & 14 & 14 & 333 & 5500 & 12900 & -61.8 & -68.0 & -64.5 & 2.0\\
45r & 136.1219 & 2.0816 & 34 & 25 & 297 & 3750 & 11400 & -46.5 & -43.1 & -40.5 & 0.6\\
45b & 136.1233 & 2.0803 & 35 & 25 & 306 & 3750 & 11400 & -46.5 & -57.3 & -50.0 & 1.9\\
46 & 136.1166 & 2.0983 & 26 & 25 & 332 & 3790 & 11400 & -50.2 & -52.6 & -50.2 & 0.5\\
47b & 136.3857 & 2.2687 & 34 & 27 & 332 & 3220 & 11000 & -42.0 & -55.0 & -46.7 & 3.5\\
47r & 136.3861 & 2.267 & 35 & 23 & 304 & 3220 & 11000 & -42.0 & -37.3 & -25.1 & 5.0\\
48b & 136.374 & 2.2628 & 29 & 21 & 332 & 3250 & 11000 & -43.2 & -51.4 & -47.0 & 1.5\\
48r & 136.3736 & 2.2615 & 29 & 22 & 332 & 3250 & 11000 & -43.2 & -39.5 & -22.2 & 8.9\\
49r & 136.4663 & 2.4678 & 29 & 23 & 290 & 3610 & 11300 & -45.7 & -42.2 & -33.0 & 2.2\\
49b & 136.4661 & 2.4693 & 31 & 23 & 292 & 3610 & 11300 & -45.7 & -52.2 & -49.1 & 0.9\\
50b & 136.5087 & 2.5108 & 31 & 25 & 332 & 3380 & 11100 & -43.5 & -48.5 & -46.5 & 0.8\\
50r & 136.5118 & 2.5083 & 28 & 23 & 10 & 3380 & 11100 & -43.5 & -40.6 & -37.5 & 1.0\\
51b & 137.058 & 2.9858 & 28 & 23 & 293 & 4350 & 11900 & -51.8 & -55.5 & -53.0 & 0.8\\
51r & 137.0567 & 2.9864 & 34 & 25 & 8 & 4350 & 11900 & -51.8 & -50.6 & -40.9 & 3.5\\
52r & 137.0662 & 2.9999 & 37 & 26 & 43 & 4390 & 12000 & -52.2 & -49.1 & -41.0 & 7.8\\
52b & 137.0683 & 3.0013 & 38 & 29 & 15 & 4390 & 12000 & -52.2 & -65.8 & -55.2 & 4.1\\
53b & 138.6143 & 1.5611 & 26 & 26 & 330 & 5450 & 13000 & -59.7 & -71.1 & -61.7 & 5.5\\
53r & 138.6158 & 1.563 & 25 & 23 & 330 & 5450 & 13000 & -59.7 & -57.6 & -54.5 & 1.3\\
54r & 136.382 & 0.8392 & 29 & 20 & 343 & 7480 & 14700 & -75.6 & -73.2 & -68.9 & 2.6\\
54b & 136.3824 & 0.838 & 20 & 17 & 332 & 7480 & 14700 & -75.6 & -83.1 & -77.9 & 2.0\\
55b & 136.7623 & 0.4548 & 27 & 16 & 343 & 5230 & 12700 & -60.9 & -65.2 & -62.7 & 5.4\\
55r & 136.7579 & 0.4522 & 24 & 18 & 343 & 5230 & 12700 & -60.9 & -59.0 & -53.2 & 6.0\\
}{\tablenotetext{a}{Galactocentric Radius}}

\Table{cccccc}{Outer Arm CO 3-2 Outflows - Derived Properties}
{{Outflow} & {Mass} & {Momentum} & {Energy} & {Dynamical} & {Momentum}\\
{Number} &  &  &  & {Age} & {Flux}\\
 & {(\msun)} & {(\msun \kms)} & {(10$^{42}$ ergs)} & {(10$^4$ years)} & {$10^{-6}$ \msun \kms yr$^{-1}$}\\}
{tab:outeroutflowsderived}
{
41r & 0.037 & 0.11 & 3.5 & 3.6 & 30.2\\
41b & 0.196 & 0.96 & 54.6 & 3.6 & 30.2\\
42r & 0.029 & 0.06 & 1.3 & 4.3 & 3.9\\
42b & 0.03 & 0.11 & 3.9 & 4.3 & 3.9\\
43r & 0.033 & 0.13 & 5.1 & 7.3 & 2.9\\
43b & 0.024 & 0.08 & 2.8 & 7.3 & 2.9\\
44r & 0.062 & 0.21 & 7.9 & 1.8 & 27.7\\
44b & 0.067 & 0.28 & 12.6 & 1.8 & 27.7\\
45r & 0.037 & 0.18 & 8.6 & 1.2 & 62.3\\
45b & 0.126 & 0.55 & 28.3 & 1.2 & 62.3\\
46 & 0.028 & 0.1 & 3.7 & - & -\\
47b & 0.187 & 1.32 & 101.0 & 0.6 & 553.0\\
47r & 0.232 & 1.84 & 164.0 & 0.6 & 553.0\\
48b & 0.054 & 0.33 & 20.8 & 0.4 & 754.0\\
48r & 0.341 & 2.4 & 229.0 & 0.4 & 754.0\\
49r & 0.106 & 0.77 & 62.7 & 1.4 & 71.1\\
49b & 0.047 & 0.22 & 10.8 & 1.4 & 71.1\\
50b & 0.037 & 0.14 & 5.5 & 3.8 & 7.4\\
50r & 0.038 & 0.14 & 5.5 & 3.8 & 7.4\\
51b & 0.058 & 0.13 & 3.0 & 1.0 & 152.0\\
51r & 0.303 & 1.33 & 72.7 & 1.0 & 152.0\\
52r & 0.829 & 4.3 & 250.0 & 1.4 & 479.0\\
52b & 0.472 & 2.5 & 150.0 & 1.4 & 479.0\\
53b & 0.626 & 3.16 & 194.0 & 4.7 & 73.2\\
53r & 0.124 & 0.31 & 8.5 & 4.7 & 73.2\\
54r & 0.461 & 1.24 & 37.5 & 1.8 & 135.0\\
54b & 0.212 & 1.14 & 64.0 & 1.8 & 135.0\\
55b & 0.411 & 1.66 & 68.0 & 7.5 & 28.3\\
55r & 0.404 & 0.47 & 6.6 & 7.5 & 28.3\\
}{
}

\appendix
\onecolumn
\section{Optically Thin, LTE dipole molecule}
\label{appendix:dipole}
While many authors have solved the problem of converting CO 1-0 beam
temperatures to \hh\ column densities
\citep{garden1991,bourke1997,Cabrit1990,lada1996}, there are no  examples in
the literature of a full derivation of the LTE, optically thin CO-to-\hh\
conversion process for higher rotational states.  We present the full
derivation here, and quantify the systematic errors generated by various
assumptions.

We begin with the assumption of an optically thin cloud such that the radiative
transfer equation \citep[][eqn 1.9]{rohlfs} simplifies to
\begin{equation}
  \label{eqn:radtrans}
  \frac{dI_\nu}{\ds} = -\kappa_\nu I_\nu 
\end{equation}

The absorption and stimulated emission terms yield 
\begin{equation}
  \label{eqn:kappa}
  \kappa_\nu = \frac{h \nu_{ul} B_{ul} n_u}{c} \varphi(\nu)
              -\frac{h \nu_{ul} B_{lu} n_l}{c} \varphi(\nu)
\end{equation}
where $\varphi(\nu)$ is the line shape function ($\int\varphi(\nu) \dnu \equiv
1$), $n$ is the density in the given state, $\nu$ is the frequency of the transition,
$B$ is the Einstein B coefficient, and $h$ is Planck's constant.

By assuming LTE (the Boltzmann distribution) and using Kirchoff's Law and the definition of 
the Einstein A and B values, we can derive a more useful version of this equation
\begin{equation}
  \kappa_\nu = \frac{c^2}{8 \pi \nu_{ul}^2} n_u A_{ul} \left[\exp\left(\frac{h \nu_{ul} }{k_B T_{ex}}\right) - 1 \right] \varphi(\nu)
\end{equation}
where $k_B$ is Boltzmann's constant.

The observable $T_B$ can be related to the optical depth, which is given by 
\begin{equation}
  \int \tau_\nu \dnu = \frac{c^2}{8 \pi \nu_{ul}^2} A_{ul} \left[\exp\left(\frac{h \nu_{ul} }{k_B T_{ex}}\right) - 1 \right] \int \varphi(\nu) \dnu \int n_u \ds 
\end{equation}

Rearranging and converting from density to column ($\int n \ds = N$) gives an equation for the column density
of the molecule in the upper energy state of the transition:
\begin{equation}
  \label{eqn:nuppertau}
  N_u = \frac{8\pi \nu_{ul}^2}{c^2 A_{ul}} \left[\exp\left(\frac{h \nu_{ul} }{k_B T_{ex}}\right) - 1 \right]^{-1} \int \tau_\nu \dnu
\end{equation}

In order to relate the brightness temperature to the optical depth, at CO transition frequencies the full blackbody
formula must be used and the CMB must also be taken into account.  \citet{rohlfs} equation 15.29 
\begin{equation} 
  \label{eqn:tbrightnesscmb}
  T_B(\nu) = \frac{h \nu}{k_B} \left(\left[e^{h \nu / k_B T_{ex}} - 1\right]^{-1} - \left[e^{h \nu / k_B T_{CMB}} - 1\right]^{-1} \right) (1-e^{-\tau_\nu})
\end{equation}
is rearranged to solve for $\tau_\nu$:
\begin{equation}
  \label{eqn:tau}
  \tau_\nu = -\ln\left[ 1 - \frac{k_B T_B}{h \nu} \left(\left[e^{h \nu / k_B T_{ex}} - 1\right]^{-1} - \left[e^{h \nu / k_B T_{CMB}} - 1\right]^{-1} \right)^{-1} \right]
\end{equation}

We convert from frequency to velocity units with $\dnu = \nu/c \dv$, and plug \eqref{eqn:tau} into \eqref{eqn:nuppertau} to get
\begin{equation}
  \label{eqn:nuppernoapprox}
  N_u = \frac{8\pi \nu_{ul}^3}{c^3 A_{ul}} \left[\exp\left(\frac{h \nu_{ul} }{k_B T_{ex}}\right) - 1 \right]^{-1} \int -\ln\left[ 1 - \frac{k_B T_B}{h \nu_{ul}} \left(\left[e^{h \nu_{ul} / k_B T_{ex}} - 1\right]^{-1} - \left[e^{h \nu_{ul} / k_B T_{CMB}} - 1\right]^{-1} \right)^{-1} \right] \dv
\end{equation}
which is the full LTE upper-level column density with no approximations applied.

The first term of the Taylor expansion is appropriate for $\tau<<1$ ($\ln[1+x]\approx x-\frac{x^2}{2}+\frac{x^3}{3}\ldots$)
\begin{equation}
  N_u = \frac{8\pi \nu_{ul}^3}{c^3 A_{ul}} \left[\exp\left(\frac{h \nu_{ul} }{k_B T_{ex}}\right) - 1 \right]^{-1} \int \frac{k_B T_B}{h \nu_{ul}} \left(\left[e^{h \nu_{ul} / k_B T_{ex}} - 1\right]^{-1} - \left[e^{h \nu_{ul} / k_B T_{CMB}} - 1\right]^{-1} \right)^{-1} \dv
\end{equation}
which simplifies to
\begin{equation}
  \label{eqn:nupper}
  N_u = \frac{8\pi \nu_{ul}^2 k_B}{c^3 A_{ul} h }  \frac{e^{h\nu_{ul}/k_B T_{CMB}} - 1}{e^{h\nu_{ul}/k_B T_{CMB}} - e^{h\nu_{ul}/k_B T_{ex}}} \int T_B  \dv
\end{equation}

This can be converted to use $\mu_e$ \citep[0.1222 for
\twelveco; ][]{Muenter1975}, the electric dipole moment of the molecule, instead
of $A_{ul}$, using \citet{rohlfs} equation 15.20 $\left((A_{ul}=(64\pi^4)/(3 h
c^3)\right)\nu^3 \mu_{e}^2$):
\begin{equation}
  \label{eqn:nuppermuju}
  N_u = \frac{3  }{8 \pi^3 \mu_e^2 } \frac{k_B}{\nu_{ul}} \frac{2 J_u + 1}{J_u} 
    \frac{e^{h\nu_{ul}/k_B T_{cmb}} - 1}{e^{h\nu_{ul}/k_B T_{CMB}} - e^{h\nu_{ul}/k_B T_{ex}}} \int T_B  \dv
\end{equation}

The total column can be derived from the column in the upper state using the partition
function and the Boltzmann distribution
\begin{equation}
  n_{tot} =        \sum_{J=0}^\infty n_J = n_0 \sum_{J=0}^\infty  (2J+1) \exp\left(-\frac{J(J+1) B_e h}{k_B T_{ex}}\right) \label{eqn:approxpartition}\\
\end{equation}
This equation is frequently approximated using an integral
\citep[e.g.][]{Cabrit1990}, but a more accurate numerical solution using up to
thousands of rotational states is easily computed
\begin{equation}
  n_J = \left[ \sum_{j=0}^{j=j_{max}} (2j+1) \exp\left(-\frac{j(j+1) B_e h}{k_B T_{ex}}\right) \right]^{-1} (2J+1) \exp\left(-\frac{J(J+1) B_e h}{k_B T_{ex}}\right)
\end{equation}
The effects of using the approximation and the full numerical solution are shown in figure \ref{fig:approx}.

\Figure{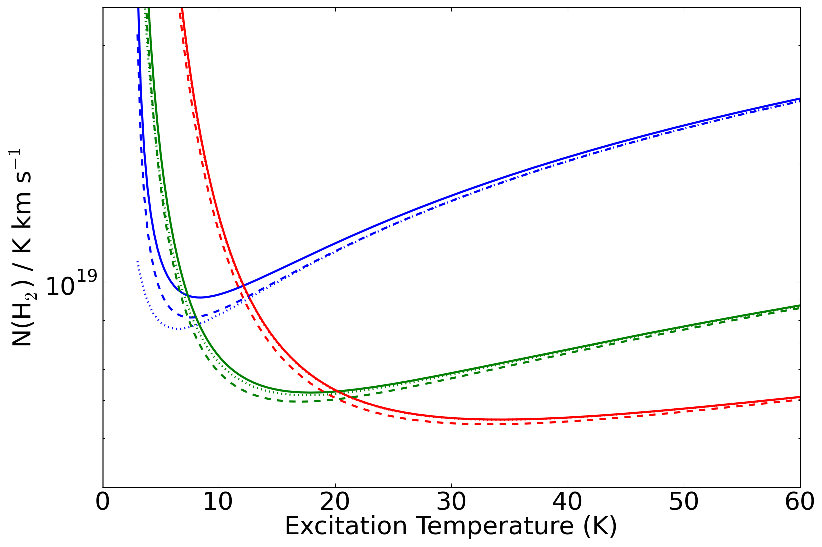}
{The LTE, optically thin conversion factor from $T_B$ (K \kms) to N(\hh)
(\persc) assuming X$_{\twelveco}=10^{-4}$ plotted against $T_{ex}$.  The
dashed line shows the effect of using the integral approximation of the 
partition function \citep[e.g.][]{Cabrit1990}.  It is a better
approximation away from the critical point, and is a better approximation
for higher transitions.  The dotted line shows the effects of removing the 
CMB term from \eqref{eqn:tbrightnesscmb}; the CMB populates the lowest two
excited states, but contributes nearly nothing to the $J=3$ state. Top (blue):
J=1-0, Middle (green): J=2-1, Bottom (red): J=3-2.}
{fig:approx}{1.0}

The CO 3-2 transition is also less likely to be in LTE than the 1-0 transition.
The critical density ($n_{cr}\equiv A_{ul}/C_{ul}$) of \twelveco\ 3-2 is 27
times higher than that for 1-0.  We have run RADEX \citep{VanDerTak2007} LVG
models of CO to examine the impact of sub-thermal excitation on column
derivation.  The results of the RADEX models are shown in Figure
\ref{fig:coradex}.  They illustrate that, while it is quite safe to assume the
CO 1-0 transition is in LTE in most circumstances, a similar assumption is
probably invalid for the CO 3-2 transition in typical molecular cloud
environments.

\Figure{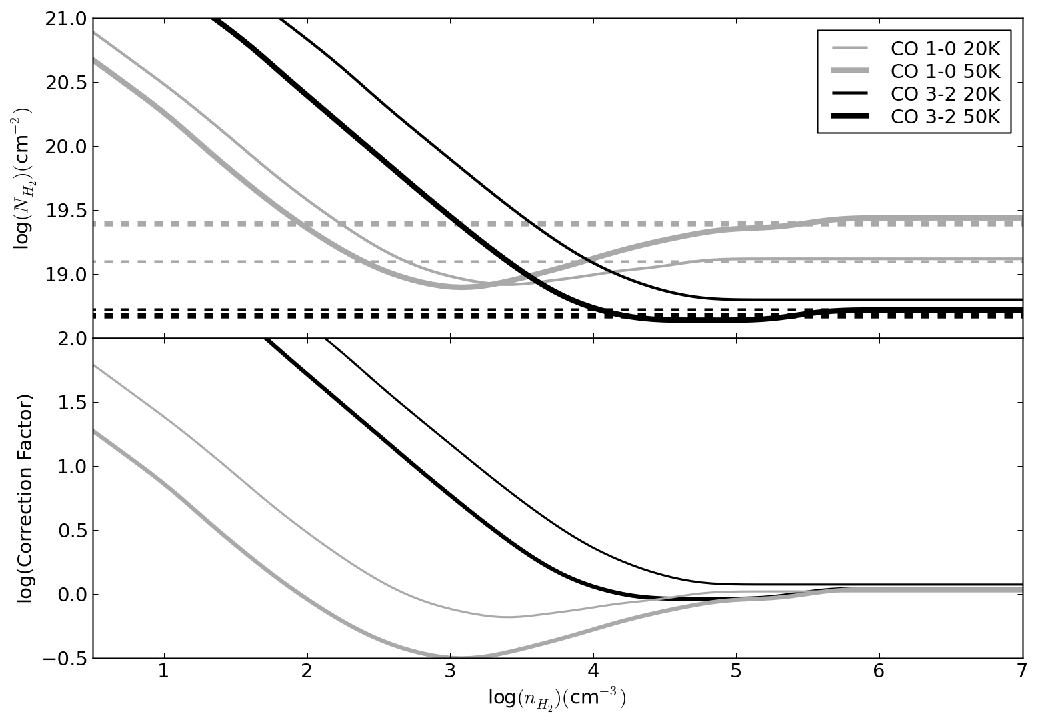}
{{\it Top}: The derived N(\hh) as a function of $n_{\hh}$ for $T_{B}=1$ K.
The dashed lines represent the LTE-derived $N(\hh)/T_B$ factor, which has 
no density dependence and, for CO 3-2, only a weak dependence on temperature.
We assume an abundance of \twelveco\ relative to \hh\ $X_{CO} = 10^{-4}$.
{\it Bottom}: The correction factor (N(\hh)$_{RADEX}$ / N(\hh)$_{LTE}$) as
a function of $n_{\hh}$.
For $T_K=20$ K, the ``correction factor'' at $10^3$ \percc\ (typical GMC
mean volume densities) is $\sim15$, while at $10^4$ \percc\ (closer to $n_{crit}$ but
perhaps substantially higher than GMC densities) it becomes negligible.  The
correction factor is also systematically lower for a higher gas kinetic
temperature.
For some densities, the ``correction factor'' dips below 1, particularly for CO
1-0.  This effect is from a slight population inversion due to fast spontaneous
decay rates from the higher levels and has been noted before
\citep[e.g.][]{Goldsmith1972}.
}{fig:coradex}{1.0}


\begin{thebibliography}{44}
\expandafter\ifx\csname natexlab\endcsname\relax\def\natexlab#1{#1}\fi

\bibitem[{{Adams} \& {Fatuzzo}(1996)}]{adams1996}
{Adams}, F.~C. \& {Fatuzzo}, M. 1996, \apj, 464, 256

\bibitem[{Arce {et~al.}(2010)Arce, Borkin, Goodman, Pineda, \&
  Halle}]{arce2010}
Arce, H.~G., Borkin, M.~A., Goodman, A.~A., Pineda, J.~E., \& Halle, M.~W.
  2010, The Astrophysical Journal, 715, 1170

\bibitem[{{Bachiller}(1996)}]{Bachiller1996}
{Bachiller}, R. 1996, \araa, 34, 111

\bibitem[{{Bally} {et~al.}(2008){Bally}, {Walawender}, {Johnstone}, {Kirk}, \&
  {Goodman}}]{bally-perseus2008}
{Bally}, J., {Walawender}, J., {Johnstone}, D., {Kirk}, H., \& {Goodman}, A.
  2008, {The Perseus Cloud}, ed. {Reipurth, B.}, 308--+

\bibitem[{{Beaumont} \& {Williams}(2010)}]{Beaumont2009}
{Beaumont}, C.~N. \& {Williams}, J.~P. 2010, \apj, 709, 791

\bibitem[{{Bertoldi} \& {McKee}(1990)}]{bertoldi:cometary:1990}
{Bertoldi}, F. \& {McKee}, C.~F. 1990, \apj, 354, 529

\bibitem[{{Bourke} {et~al.}(1997){Bourke}, {Garay}, {Lehtinen}, {Koehnenkamp},
  {Launhardt}, {Nyman}, {May}, {Robinson}, \& {Hyland}}]{bourke1997}
{Bourke}, T.~L. {et~al.} 1997, \apj, 476, 781

\bibitem[{Bretherton {et~al.}(2002)Bretherton, Moore, \&
  Ridge}]{bretherton:unbiased:2002}
Bretherton, D.~E., Moore, T. J.~T., \& Ridge, N.~A. 2002, in Hot Star Workshop
  III: The Earliest Phases of Massive Star Birth, Vol. 267, 347

\bibitem[{{Cabrit} \& {Bertout}(1990)}]{Cabrit1990}
{Cabrit}, S. \& {Bertout}, C. 1990, \apj, 348, 530

\bibitem[{{Chernin} \& {Masson}(1991)}]{Chernin1991}
{Chernin}, L.~M. \& {Masson}, C.~R. 1991, \apjl, 382, L93

\bibitem[{{Curtis} {et~al.}(2010){Curtis}, {Richer}, {Swift}, \&
  {Williams}}]{curtis2010}
{Curtis}, E.~I., {Richer}, J.~S., {Swift}, J.~J., \& {Williams}, J.~P. 2010,
  \mnras, 1304

\bibitem[{{Cyganowski} {et~al.}(2009){Cyganowski}, {Brogan}, {Hunter}, \&
  {Churchwell}}]{Cyganowski2009}
{Cyganowski}, C.~J., {Brogan}, C.~L., {Hunter}, T.~R., \& {Churchwell}, E.
  2009, \apj, 702, 1615

\bibitem[{{Deharveng} {et~al.}(1997){Deharveng}, {Zavagno}, {Cruz-Gonzalez},
  {Salas}, {Caplan}, \& {Carrasco}}]{Deharveng1997}
{Deharveng}, L., {Zavagno}, A., {Cruz-Gonzalez}, I., {Salas}, L., {Caplan}, J.,
  \& {Carrasco}, L. 1997, \aap, 317, 459

\bibitem[{{Digel} {et~al.}(1996){Digel}, {Lyder}, {Philbrick}, {Puche}, \&
  {Thaddeus}}]{Digel1996}
{Digel}, S.~W., {Lyder}, D.~A., {Philbrick}, A.~J., {Puche}, D., \& {Thaddeus},
  P. 1996, \apj, 458, 561

\bibitem[{{Elmegreen}(1998)}]{elmegreen1998}
{Elmegreen}, B.~G. 1998, in Astronomical Society of the Pacific Conference
  Series, Vol. 148, Origins, ed. {C.~E.~Woodward, J.~M.~Shull, \&
  H.~A.~Thronson Jr.}, 150--+

\bibitem[{{Elmegreen} \& {Lada}(1977)}]{elmegreen:sequential:1977}
{Elmegreen}, B.~G. \& {Lada}, C.~J. 1977, \apj, 214, 725

\bibitem[{{Enoch} {et~al.}(2006){Enoch}, {Young}, {Glenn}, {Evans}, {Golwala},
  {Sargent}, {Harvey}, {Aguirre}, {Goldin}, {Haig}, {Huard}, {Lange},
  {Laurent}, {Maloney}, {Mauskopf}, {Rossinot}, \& {Sayers}}]{Enoch2006}
{Enoch}, M.~L. {et~al.} 2006, \apj, 638, 293

\bibitem[{{Evans} {et~al.}(2009){Evans}, {Dunham}, {J{\o}rgensen}, {Enoch},
  {Mer{\'{\i}}n}, {van Dishoeck}, {Alcal{\'a}}, {Myers}, {Stapelfeldt},
  {Huard}, {Allen}, {Harvey}, {van Kempen}, {Blake}, {Koerner}, {Mundy},
  {Padgett}, \& {Sargent}}]{Evans2009}
{Evans}, N.~J. {et~al.} 2009, \apjs, 181, 321

\bibitem[{{Felli} {et~al.}(1987){Felli}, {Hjellming}, \&
  {Cesaroni}}]{Felli1987}
{Felli}, M., {Hjellming}, R.~M., \& {Cesaroni}, R. 1987, \aap, 182, 313

\bibitem[{{Garden} {et~al.}(1991){Garden}, {Hayashi}, {Hasegawa}, {Gatley}, \&
  {Kaifu}}]{garden1991}
{Garden}, R.~P., {Hayashi}, M., {Hasegawa}, T., {Gatley}, I., \& {Kaifu}, N.
  1991, \apj, 374, 540

\bibitem[{{Ginsburg} {et~al.}(2009){Ginsburg}, {Bally}, {Yan}, \&
  {Williams}}]{Ginsburg2009}
{Ginsburg}, A.~G., {Bally}, J., {Yan}, C., \& {Williams}, J.~P. 2009, \apj,
  707, 310

\bibitem[{{Goldsmith}(1972)}]{Goldsmith1972}
{Goldsmith}, P.~F. 1972, \apj, 176, 597

\bibitem[{{Gritschneder} {et~al.}(2010){Gritschneder}, {Burkert}, {Naab}, \&
  {Walch}}]{Gritschneder2010}
{Gritschneder}, M., {Burkert}, A., {Naab}, T., \& {Walch}, S. 2010, \apj, 723,
  971

\bibitem[{{Hachisuka} {et~al.}(2006){Hachisuka}, {Brunthaler}, {Menten},
  {Reid}, {Imai}, {Hagiwara}, {Miyoshi}, {Horiuchi}, \&
  {Sasao}}]{Hachisuka2006}
{Hachisuka}, K. {et~al.} 2006, \apj, 645, 337

\bibitem[{Hatchell \& Dunham(2009)}]{hatchell2009}
Hatchell, J. \& Dunham, M.~M. 2009, 0904.1163

\bibitem[{Hatchell {et~al.}(2007)Hatchell, Fuller, \& Richer}]{hatchell2007}
Hatchell, J., Fuller, G.~A., \& Richer, J.~S. 2007, Astronomy and Astrophysics,
  472, 187

\bibitem[{{Heyer} {et~al.}(1998){Heyer}, {Brunt}, {Snell}, {Howe}, {Schloerb},
  \& {Carpenter}}]{heyer:ogs:1998}
{Heyer}, M.~H., {Brunt}, C., {Snell}, R.~L., {Howe}, J.~E., {Schloerb}, F.~P.,
  \& {Carpenter}, J.~M. 1998, \apjs, 115, 241

\bibitem[{Karr \& Martin(2003)}]{karr:triggered:2003}
Karr, J.~L. \& Martin, P.~G. 2003, Astrophysical Journal, 595, 900

\bibitem[{{Klein} {et~al.}(1983){Klein}, {Sandford}, \&
  {Whitaker}}]{klein:implosion:1983}
{Klein}, R.~I., {Sandford}, II, M.~T., \& {Whitaker}, R.~W. 1983, \apjl, 271,
  L69

\bibitem[{Koenig {et~al.}(2008)Koenig, Allen, Gutermuth, Hora, Brunt, \&
  Muzerolle}]{koenig:clustered:2008}
Koenig, X.~P., Allen, L.~E., Gutermuth, R.~A., Hora, J.~L., Brunt, C.~M., \&
  Muzerolle, J. 2008, Astrophysical Journal, 688, 1142

\bibitem[{{Lada} \& {Fich}(1996)}]{lada1996}
{Lada}, C.~J. \& {Fich}, M. 1996, \apj, 459, 638

\bibitem[{Lefloch {et~al.}(1997)Lefloch, Lazareff, \&
  Castets}]{lefloch:cometary:1997}
Lefloch, B., Lazareff, B., \& Castets, A. 1997, Astronomy and Astrophysics,
  324, 249

\bibitem[{{Muenter}(1975)}]{Muenter1975}
{Muenter}, J. 1975, Journal of Molecular Spectroscopy, 55, 490, is always cited
  for $\mu=0.122$ debye.

\bibitem[{Oey {et~al.}(2005)Oey, Watson, Kern, \&
  Walth}]{oey:hierarchical:2005}
Oey, M.~S., Watson, A.~M., Kern, K., \& Walth, G.~L. 2005, Astronomical
  Journal, 129, 393

\bibitem[{Peters {et~al.}(2010)Peters, Banerjee, Klessen, Low,
  {Galván-Madrid}, \& Keto}]{peters2010}
Peters, T., Banerjee, R., Klessen, R.~S., Low, M.~M., {Galván-Madrid}, R., \&
  Keto, E.~R. 2010, The Astrophysical Journal, 711, 1017

\bibitem[{{Reid} {et~al.}(2009){Reid}, {Menten}, {Zheng}, {Brunthaler},
  {Moscadelli}, {Xu}, {Zhang}, {Sato}, {Honma}, {Hirota}, {Hachisuka}, {Choi},
  {Moellenbrock}, \& {Bartkiewicz}}]{Reid2009}
{Reid}, M.~J. {et~al.} 2009, ArXiv e-prints

\bibitem[{Reipurth \& Bally(2001)}]{reipurth2001}
Reipurth, B. \& Bally, J. 2001, Annual Review of Astronomy and Astrophysics,
  39, 403

\bibitem[{{Shu} {et~al.}(1987){Shu}, {Adams}, \& {Lizano}}]{Shu1987}
{Shu}, F.~H., {Adams}, F.~C., \& {Lizano}, S. 1987, \araa, 25, 23

\bibitem[{{Snell} {et~al.}(2002){Snell}, {Carpenter}, \& {Heyer}}]{Snell2002}
{Snell}, R.~L., {Carpenter}, J.~M., \& {Heyer}, M.~H. 2002, \apj, 578, 229

\bibitem[{{Stanke} {et~al.}(1999){Stanke}, {McCaughrean}, \&
  {Zinnecker}}]{Stanke1999}
{Stanke}, T., {McCaughrean}, M.~J., \& {Zinnecker}, H. 1999, \aap, 350, L43

\bibitem[{{Taylor} {et~al.}(2003){Taylor}, {Gibson}, {Peracaula}, {Martin},
  {Landecker}, {Brunt}, {Dewdney}, {Dougherty}, {Gray}, {Higgs}, {Kerton},
  {Knee}, {Kothes}, {Purton}, {Uyaniker}, {Wallace}, {Willis}, \&
  {Durand}}]{Taylor2003:CGPS}
{Taylor}, A.~R. {et~al.} 2003, AJ, 125, 3145

\bibitem[{Thompson {et~al.}(2004)Thompson, White, Morgan, Miao, Fridlund, \&
  {Huldtgren-White}}]{thompson:searching:2004}
Thompson, M.~A., White, G.~J., Morgan, L.~K., Miao, J., Fridlund, C. V.~M., \&
  {Huldtgren-White}, M. 2004, Astronomy and Astrophysics, 414, 1017

\bibitem[{{van der Tak} {et~al.}(2007){van der Tak}, {Black}, {Sch{\"o}ier},
  {Jansen}, \& {van Dishoeck}}]{VanDerTak2007}
{van der Tak}, F.~F.~S., {Black}, J.~H., {Sch{\"o}ier}, F.~L., {Jansen}, D.~J.,
  \& {van Dishoeck}, E.~F. 2007, \aap, 468, 627

\bibitem[{Wilson {et~al.}(2009)Wilson, Rohlfs, \& H\"{u}ttemeister}]{rohlfs}
Wilson, T.~L., Rohlfs, K., \& H\"{u}ttemeister, S. 2009, Tools of radio
  astronomy (Springer)

\end{thebibliography}
\end{document}